\PassOptionsToPackage{
  pdfauthor={Inge Madshaven et al.},
  pdftitle={%
    Simulation model for the propagation of second mode streamers
  	in dielectric liquids using the Townsend--Meek criterion%
    },
}{hyperref}

\documentclass[10pt]{iopart}
\usepackage{iopams}

\expandafter\let\csname equation*\endcsname\relax
\expandafter\let\csname endequation*\endcsname\relax
\newcommand{\newblock}{}  

\usepackage{etoolbox}                   
\usepackage{amsmath}   

\usepackage[utf8]{inputenc}
\usepackage{lmodern}    
\usepackage{bm}

\usepackage{graphics}
\usepackage{epstopdf}     
\usepackage{booktabs}       
\usepackage{tabularx}
\usepackage{url} 
\usepackage{tikz}

\usepackage[detect-all,per-mode=reciprocal]{siunitx}        

\usepackage[square,numbers,sort&compress]{natbib}
\usepackage[%
  ocgcolorlinks,%
  linktoc=page,%
  breaklinks,%
  hyperfootnotes=false,%
  hidelinks,
  ]{hyperref}

\usepackage{cleveref}
\crefname{equation}{}{}
\Crefname{equation}{Equation}{Equations}
\crefname{figure}{figure}{figures}
\Crefname{figure}{Figure}{Figures}
\crefname{table}{table}{tables}
\Crefname{table}{Table}{Tables}
\crefname{appendix}{}{}
\Crefname{appendix}{}{}

\renewcommand{\vec}[1]{\bm{#1}}
\let\oldhat\hat
\renewcommand{\hat}[1]{\oldhat{\bm{#1}}}

\usetikzlibrary{shapes.geometric, arrows.meta}
\tikzset{%
  >={Latex[width=2mm,length=2mm]},
            base/.style = {rectangle, rounded corners, draw=black,
                           minimum width=18 ex, minimum height=5 ex,
                           inner sep=2 ex,
                           text centered, font=\sffamily,
                           },
  activityStarts/.style = {base, fill=blue!30},
       startstop/.style = {base, fill=red!30},
    activityRuns/.style = {base, fill=green!30},
         process/.style = {base, fill=orange!15},
        decision/.style = {base, diamond, aspect=2, inner sep=1ex, fill=yellow!30},
}

\makeatletter

\begingroup
\catcode`_=\active
\catcode`^=\active
\protected\gdef_{\@ifnextchar'\subtextrm{\@ifnextchar"\subtextsc\sb}}
\protected\gdef^{\@ifnextchar'\suptextrm{\@ifnextchar"\suptextsc\sp}}
\endgroup
\def\subtextrm'#1'{\sb{\textrm{#1}}}
\def\suptextrm'#1'{\sp{\textrm{#1}}}
\def\subtextsc"#1"{\sb{\textsc{#1}}}
\def\suptextsc"#1"{\sp{\textsc{#1}}}
\AtBeginDocument{\catcode`_=12 \mathcode`_="8000}
\AtBeginDocument{\catcode`^=12 \mathcode`^="8000}

\newcommand{\abs}[1]{{\lvert#1\rvert}}

\renewcommand{\d}[1]{\text{d}{#1}}

\makeatother

\begin{document}



\title[Simulation model for the propagation of second mode streamers]{%
  Simulation model for the propagation of second mode streamers
  in dielectric liquids using the Townsend--Meek criterion%
  }


\author{
  I.~Madshaven$^1$,
  P.-O.~Åstrand$^1$%
    \footnote{Corresponding author: \texttt{per-olof.aastrand@ntnu.no}},
  O. L.~Hestad$^2$,
  S.~Ingebrigtsen$^2$%
    \footnote{Present address: ABB AS, 5257 Kokstad, Norway},
  M.~Unge$^3$,
  O.~Hjortstam$^3$
  }

\address{%
  $^1$
  Department of Chemistry,
  NTNU, Norwegian University of Science and Technology,
  7491 Trondheim, Norway\\%
}
\address{%
  $^2$
  SINTEF Energy Research,
  7465 Trondheim, Norway%
}
\address{%
  $^3$
  ABB Corporate Research,
  72178 Västerås, Sweden%
}


\begin{abstract}
A simulation model for
second mode positive streamers
in dielectric liquids is presented.
Initiation and propagation
is modeled
by an electron-avalanche mechanism
and the Townsend--Meek criterion.
The electric breakdown is simulated
in a point-plane gap,
using cyclohexane as a model liquid.
Electrons move
in a Laplacian electric field
arising from the electrodes and streamer structure,
and turn into electron avalanches in high-field regions.
The Townsend--Meek criterion determines when
an avalanche is regarded as a part of the streamer structure.
The results show that an avalanche-driven breakdown is possible,
however,
the inception voltage is relatively high.
Parameter variations are included to investigate
how the parameter values affect the model.%
%
%
\end{abstract}

%
\vspace{2pc}
\noindent{\it Keywords}:
Simulation model,
Streamer,
Electron avalanche,
Townsend--Meek criterion,
Dielectric liquid,
Electrical breakdown
%

\ioptwocol
%


\section{Introduction to streamers}\label{sec:introduction}{

Dielectric liquids are widely used for insulation of high power equipment,
such as transformers,
since
liquid insulation has good cooling properties,
high electrical withstand strength,
and recovers from an electrical discharge within short time%
~\citep{Wedin2014}.
Electric breakdown in liquids is preceded by the formation of
a prebreakdown channel called a streamer~\citep{Lesaint2016}.
A partial discharge,
a local electric breakdown,
changes the electric field distribution,
which could cause another local breakdown,
and in this way, a streamer may propagate through a liquid.
A streamer bridging the gap between two electrodes,
for instance an energized part and a grounded part,
lowers the electrical withstand strength
and may cause a complete electric breakdown,
possibly destroying the equipment%
~\citep{Wedin2014}.

A streamer consists of a gaseous and partly ionized structure,
originating in one location
and branching out in filaments as it propagates through the liquid.
This structure may be observed through shadowgraphic or schlieren photography
since its refractive index differs from the surrounding liquid%
~\citep{Devins1981}.
Streamers are classified as positive or negative,
depending on the polarity of the initiation site.
Streamer experiments are often carried out in needle-plane gaps
since a strongly divergent field
allows control of where the streamer initiates,
the polarity of the streamer,
and also enables the study of streamers
that initiate, propagate, and then stops without causing an electric breakdown%
~\citep{Devins1981,Lesaint2016}.
Conversely, in a gap with a uniform field,
inception governs the breakdown probability,
since an initiated streamer
is always able to propagate the gap
due to the high background field.

The nature of streamers has been investigated for decades%
~\citep{Devins1981,Felici1988,Gournay1993,Tobazeon1994,Biller1996,Massala1998,Lundgaard1998,Kolb2008,Ingebrigtsen2009,Joshi2009,Smalo2011,Joshi2013-I,Wedin2014,Lesaint2016},
but is still not well understood.
For positive streamers in non-polar liquids,
it is common to define four distinct modes of propagation,
mainly characterized by their speed%
~\citep{Lesaint1998,Lesaint2016}.
The streamer mode depends on the applied voltage,
and may change during propagation.
The 1st mode
propagates in a bubbly or bushy fashion
with a speed
of the order of \SI{100}{\metre\per\second},
the 2nd mode is faster,
of the order of \SI{1}{\kilo\metre\per\second},
and has a branched or tree-like structure.
The even faster 3rd and 4th modes propagates at speeds
of the order of
\SI{10}{\kilo\metre\per\second} and
\SI{100}{\kilo\metre\per\second}, respectively.
The 1st mode is only observed for very sharp needles
and will usually not lead to a breakdown by itself,
but the streamer may change to the 2nd mode.
The 2nd mode may initiate for voltages below the voltage required for breakdown,
and increases in propagation length and number of branches at higher voltages.
Often, a 2nd mode streamer sporadically emits visible light~\citep{Devins1981}, re-illuminations,
from one or more of its branches.
Above the breakdown voltage,
streamers may change between the 2nd, the 3rd, and the 4th mode during propagation.
There are usually more re-illuminations in the 3rd mode than the 2nd mode.
The inception of the 4th mode is associated
with a drastic increase in speed and fewer, more luminous, branches~\citep{Lesaint2016}.

There are numerous mechanisms that can be involved in the streamer phenomena,
the challenge is identifying their importance during initiation and propagation.
Applying a potential to a needle can cause charge injection,
giving a
space-charge limited current~\citep{Dumitrescu2001}
causing
Joule heating~\citep{Dumitrescu2001},
which in turn can cause
bubble nucleation~\citep{Kattan1989}.
A breakdown in the gas bubble can then propagate the needle potential,
and the process may repeat.
This is one way to explain 1st mode propagation.
Electric fields can also cause electrohydrodynamic flow,
which could cause streamer formation through cavitation~\citep{Tobazeon1994pre}.
Electrostatic cracking has also been proposed as a cavitation mechanism~\citep{Lewis1998}.
A main topic of discussion is whether
a lowering of the liquid density is needed before charge generation can occur.
Electron avalanches are important in gas discharge,
but their importance in liquid breakdown is still disputed.
In water,
strong scattering could prevents electrons
from forming avalanches in the liquid phase~\citep{Joshi2004}.
Therefore,
discharges in micro-bubbles can be important for charge generation%
~\citep{Joshi2004,Kolb2008,Joshi2013-I}.
The same mechanism was also proposed for non-polar liquids~\citep{Lewis1998},
however, the relative permittivity is about 80 in water
and about 2 in a typical oil,
and this difference can prove important
since the field enhancement within a bubble in oil is much lower than in water.
Contrary to water,
there are indications of electron avalanches in non-polar liquids%
~\citep{Derenzo1974,Haidara1991,Dumitrescu2001},
furthermore,
while the initiation
and the propagation length of 2nd mode streamers are dependent on the pressure,
their propagation velocity is not pressure dependent~\citep{Lesaint1994,Dumitrescu2001}.
This implies that the mechanism responsible for propagation occurs in the liquid phase
and that the gaseous channel follows as a consequence.
In very high electric fields,
field-ionization can occur~\citep{Halpern1969a,Denat1988},
and this mechanism has been proposed for the fast 3rd and 4th propagation modes%
\cite{Biller1996}.
As the streamer gains length, the properties of the channel could also prove important.
The streamer channel is a partly ionized, low-temperature plasma,
having a varying conductance%
~\citep{Atten1993,Massala1998}.
The mechanisms involved when a plasma is in contact with a liquid is often overlooked
and is in itself a very complex problem%
~\cite{Bruggeman2016}.

The development of models is important
for improving electrical equipment
as well as the prevention of equipment failure.
An early simulation model for liquid breakdown
uses a lattice to
investigate the fractal nature
of the streamer structure
as a function of the electric field $E$~\citep{Niemeyer1984},
and has been expanded
to incorporate
needle-plane geometry~\citep{Wiesmann1986},
a 3D-lattice~\citep{Satpathy1986},
statistical time~\citep{Biller1993},
availability of seed electrons~\citep{Kim2008},
and
varying conductance of the streamer channels~\citep{Kupershtokh2006}.
Charge generation and transport in an electric field
have also been solved by a
finite element method (FEM) approach,
to simulate streamer propagation
in 2D and 3D,
adding impurities to generate streamer branching%
~\citep{Qian2005,Qian2006,Hwang2012,Jadidian2013}.
A major difference between breakdown in gases and liquids
is that a phase change is involved
when making the streamer channel in liquids.
The phase change is difficult to model,
but it is possible to make approximations~\citep{Naidis2015jpd},
or to focus on the plasma within the channel~\citep{Naidis2016}.

Both lattice and FEM simulations
require considerable computational power,
and therefore, the simulations are often
done for either very short timescales
or very simplified models.
The work presented here is based on~\citep{Hestad2014},
which
chooses a different approach.
It is a computational model
for 2nd mode positive streamers
in non-polar liquids,
driven by electron avalanches in the liquid phase.
A point-plane geometry is modeled,
with the point being a positively charged hyperbolic needle.
Cyclohexane is used as a model liquid,
since it is a well defined system used extensively in experiments%
~\citep{Denat1988,Haidara1991,Gournay1993,Lesaint2000,Ingebrigtsen2009}.

The model and the theoretical background is presented in \cref{sec:model},
as well as the parameters and the algorithm used for the simulations.
In \cref{sec:results}, the results are given and discussed.
First a baseline is established,
then parameter variations
and alternative parameter values are investigated.
A general discussion,
outlining the weaknesses and strengths of the model,
is given in \cref{sec:discussion}.
Finally, the main conclusions are summarized in \cref{sec:conclusion}.
\Cref{sec:pscoords}
contains additional details on
the coordinate system used in the model.

}

%


\section{Simulation model and theory}{\label{sec:model}

The model is built on the assumption
that electron avalanches occur in the liquid phase,
and that these govern the propagation of 2nd mode, positive streamers%
~\citep{Hestad2014}.
Applying a potential to the needle
in a needle-plane geometry
gives rise to an electric field.
A number of anions and electrons,
assumed to be already present in the liquid,
are accelerated by the electric field.
Subsequently,
electron multiplication occurs
in areas where the electric field is sufficiently strong,
turning electrons into electron avalanches.
An avalanche is assumed to be ``critical''
if it reaches a magnitude
given by the Townsend--Meek criterion~\citep{Pedersen1989},
and the position of such an avalanche
is regarded as a part of the streamer.
Then the electric field is reevaluated,
accounting for the potential of both the needle and the streamer.
This work investigates liquid cyclohexane as the insulating liquid,
with the option to add dimethylaniline (DMA) as an additive,
but the model can be used
for other base liquids and additives as well,
if the parameter values are available.


\subsection{Geometrical and electrical properties}

\begin{figure}[t]
    \centering
    \includegraphics[width=1.0\linewidth]{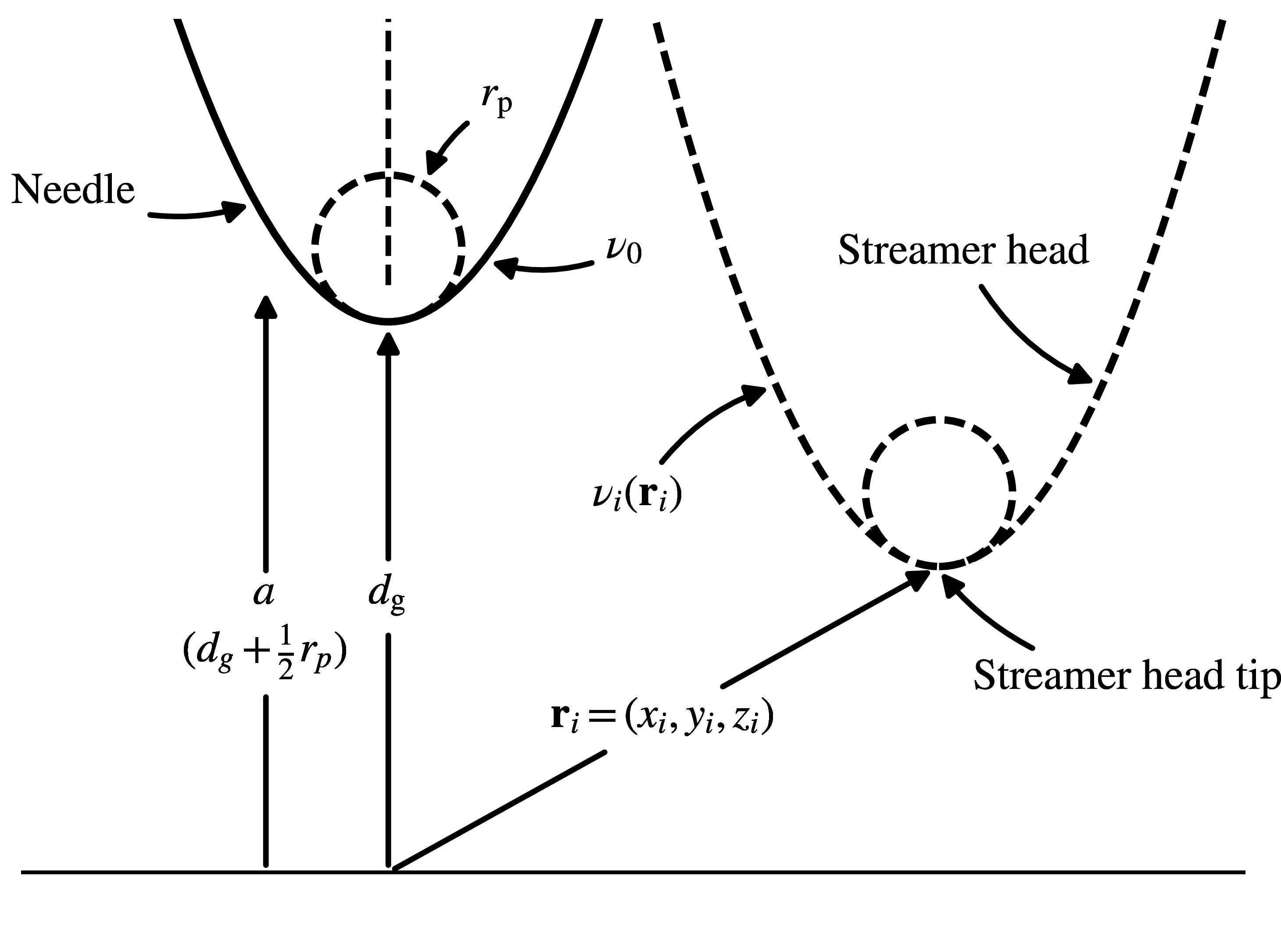}
    \caption{The hyperbolic needle and a streamer head,
            with relevant variables shown.
            The distance to the plane is usually far greater
            than illustrated here.
            }
    \label{fig:geometry}
\end{figure}

A hyperbolic needle electrode with a tip radius $r_'p'$
is placed at a distance $d_'g'$ from a planar electrode,
as illustrated in \cref{fig:geometry}
where all important geometric variables are shown.
In prolate spheroid coordinates ($\mu, \nu, \phi; a$),
a hyperboloid is represented by a single coordinate $\nu$,
and the 3D Laplace equation becomes separable,
see \cref{sec:pscoords} for details and definitions.
The potential is (cf.~\cref{eq:ps_V})
\begin{equation}
    V_i = C_i \ln \tan \frac{\nu_i}{2} \,,
    \label{eq:ps_Vi}
\end{equation}
and the electric field
is (cf.~\cref{eq:ps_E})
\begin{equation}
    \vec{E}_i
        = \frac{C_i \, \hat{\nu}_i}{h_{\nu_i} \sin \nu_i}  \,,
    \label{eq:ps_Ei}
\end{equation}
where $C_i$ is a constant.
The subscript $i$ refers to a given hyperboloid
(the needle or a streamer head),
hence,
the subscript in $\nu_i$ implies a transformation to a coordinate system
centered at hyperboloid $i$,
\begin{align}
    \nu_i (\vec{r})
    = \nu \big( x - x_i, y - y_i, z ; a_i  \big) \,.
\end{align}
The constant $C_i$ (cf.~\cref{eq:ps_C})
is given by the boundary condition,
the potential at the surface $\nu_i(\vec{r}_i)$,
\begin{equation}
    C_i
      \approx \frac{2 V_i (\vec{r}_i) }{\ln \left( 4 z_i / r_{\mathrm{p},i} \right)} \,,
    \label{eq:ps_Ci}
\end{equation}
which is valid for a sharp needle,
$r_'p' \ll z_i$.
The other boundary condition,
that the potential is zero at the plane
$V_i(\vec{r}\hat{z} = 0) = 0$,
is already accounted for.
For the needle,
$V_i(\vec{r}_i) = V_0$,
which is the applied potential.
Calculating the electric field in~\cref{eq:ps_Ei}
is the most expensive part of the computer simulation,
although explicit calculation
of the trigonometric functions
can be avoided (cf. \cref{sec:pscoords}).
Using the Laplace equation instead of the Poisson equation
is a simplification that will be discussed further in \cref{sec:discussion}.

%


\subsection{Electrons and ions in dielectric liquids}\label{sec:elecronsandions}

Naturally occurring radiation is of the order of
$D_'r' = \SI{1}{mSv}$ per year~\citep{UNSCEAR2008} and may produce electron-cation pairs
by ionizing neutral molecules.
The production rate is~\citep{Schmidt1997}
\begin{equation}
    R_'e' = D_'r' \, \rho \, G \,,
    \label{eq:R_e}
\end{equation}
where
the density $\rho$ is $\SI{0.78}{kg \per l}$ for cyclohexane.
The yield $G$ is usually given in events per \SI{100}{eV}.
Hydrocarbons typically have an ion yield $G_'ion'$ of about 4~\citep{Holroyd2003},
and for cyclohexane it is 4.3~\citep{Gee1992}.
However, the free electron yield $G_'free'$ is much lower,
about \SI{0.15}{}~\citep{Dodelet1972,Gee1992},
which implies that most electrons recombines geminately.
This gives a production of $R_'e' = \SI{2.3e8}{m^{-3} s^{-1}}$.
The recombination process is rapid,
and the electron lifetime is~\citep{Schmidt1997}
\begin{equation}
    \tau_'r' = \frac{4 \pi \epsilon_0 \epsilon_'r' r_0^3}{3 \mu_'el' e} \,,
\end{equation}
where
$\epsilon_0$ is the vacuum permittivity,
$\epsilon_'r' = 2.0$ is the typical relative permittivity for hydrocarbons,
$r_0$ is the recombination distance,
$\mu_'e'$ is the electron mobility,
and
$e$ is the elementary charge.
Inserting the thermalization distance (the most likely distance)
$r_0 = \SI{5.9}{nm}$\citep{Gee1992} and
a mobility $\mu_'e' = \SI{45}{mm^2 V^{-1} s^{-1}}$~\citep{Schmidt1970,Dodelet1972},
yields $\tau_'r' = \SI{1.7}{ps}$.

The average drift velocity $v_'d'$ of an electron or ion
is given by its mobility $\mu$ and the local electric field $\vec{E}$,
\begin{equation}
    \vec{v}_'d' = \mu \vec{E} \,.
    \label{eq:v_d}
\end{equation}
In liquids where the electron mobility is low ($\mu_'e' < \SI{e2}{mm^2 V^{-1} s^{-1}}$),
the electron is regarded as localized,
and electron transport is explained
either through a hopping or a trapping mechanism~\citep{Schmidt1984,Holroyd1989}.
The drift velocity is proportional to the electric field when the electric strength is low,
however,
for low-mobility liquids,
it becomes superlinear in high fields~\citep{Schmidt1984,Schmidt1997}.
The lifetimes of free electrons and ions
can be related to the reaction rates.
The reaction rate constants $k_'r'$
are found by the Debye relation~\citep{Winokur1975,Schmidt1997},
\begin{equation}
    k_'r' = \frac{e}{\epsilon_0 \epsilon_'r'} (\mu_- + \mu_+) \,,
    \label{eq:k_r}
\end{equation}
where
$\mu_\pm$ is the mobility of the respective reacting species.
This relation assumes that recombination is limited by diffusion,
which is related to the mobilities,
and the relation holds as long as the mobilities are low
($<$ \SI{e4}{mm^2 V^{-1} s^{-1}})~\citep{Schmidt1997}.
In cyclohexane,
the ion mobility is of the order of
\SI{e-2}{} to
\SI{e-1}{mm^2 V^{-1} s^{-1}}%
~\citep{Polak1977,Denat1979,Alj1985,Denat1988,Gee1992,Dumitrescu2001}
and the electron mobility is of the order of \SI{10}{mm^2 V^{-1} s^{-1}}~\citep{Dodelet1972,Allen1976,Schmidt1977,Gee1992}.
Using
$\mu_'e' = \SI{45}{mm^2 V^{-1} s^{-1}}$
and
$\mu_'ion' = \SI{0.1}{mm^2 V^{-1} s^{-1}}$,
yields
$k_'r' = \SI{4.1e-13}{m^{3} \per s}$ for electron-ion recombination and
$k_'r' = \SI{1.8e-15}{m^{3} \per s}$ for ion-ion recombination
according to \cref{eq:k_r}.
This implies that there is a far greater number of anions than electrons.
However, small impurities, such as $\text{O}_2$,
have higher mobilities~\citep{Schmidt1997}.

The low-field conductivity for the liquid $\sigma$ is given by
the number density of charge carriers $n_i$ for species $i$
and their mobilities,
\begin{equation}
    \sigma = e \sum_i n_i \mu_i \,.
    \label{eq:sigma}
\end{equation}
By assuming that the measured conductivity is due to ions only
and that the ions are similar in number and mobility,
the number density of the anions is
\begin{equation}
    n_'ion' = \frac{\sigma} {2 e \mu_'ion'} \,,
    \label{eq:n_ion}
\end{equation}
which yields $n_'ion' = \SI{6.2e12}{m^{-3}}$
for $\sigma = \SI{0.2}{pS \per m}$~\citep{Alj1985,Denat2011}.
A similar result is obtained by considering a steady-state condition,
\begin{equation}
    \frac{\d n_'e'}{\d t}
        = R_'e' - k_'r' n_'e' n_'p' - \frac{n_'e'}{\tau_'a'} = 0\,,
    \label{eq:dne}
\end{equation}
where
$n_'e'$ is the electron density,
$n_'p'$ is the cation density,
and
$t$ is the time.
If the electron attachment time $\tau_'a'$ is large~\citep{Kim2007},
\begin{equation}
    n_'e' \approx n_'p' \approx \sqrt{\frac{R_'e'}{k_'r'}} \,,
    \label{eq:n_ep}
\end{equation}
which yields
$n_'e' = \SI{2.4e10}{m^{-3}}$.
However, $\tau_'a'$ is assumed small,
about \SI{200}{\nano\second}~\citep{Jadidian2013},
which implies that $n_'ion' \approx n_'p'$.
Using \cref{eq:n_ep}
with the ion-ion recombination rate
yields
$n_'ion' = \SI{3.6e11}{m^{-3}}$,
about an order of magnitude
lower than what obtained from \cref{eq:n_ion}.
With rapid attachment,
\cref{eq:dne} is
\begin{equation}
    n_'e' \approx R_'e' \, \tau_'a' \,.
\end{equation}
and yields
$n_'e' = \SI{46}{m^{-3}}$,
which shows that the assumption $n_'ion' \approx n_'p'$ holds.

%


\subsection{Electron avalanches}

The main concept the model is that electrical breakdown
is driven by electron avalanches
occurring in the liquid phase~\citep{Haidara1991,Ingebrigtsen2009,Hestad2014}.
A number of anions, calculated by \cref{eq:n_ion},
is considered as the source of electrons
by an electron-detachment mechanism.
These electrons initiates the avalanches.
As shown in \cref{sec:elecronsandions}, the number of anions is far greater than the number of electrons,
and it is also far greater than the number of electrons produced
within a simulation
(a volume less than \SI{1}{cm^3} and a time less than \SI{1}{s}).

The needle electrode and the streamer creates an electric field $\vec{E}$.
Transformer oils experience increased conductivity
due to ion dissociation
when the electric field exceeds some \si{MV\per m}~\citep{Gafvert1992}.
The model assumes that also electrons detach from anions
for field strengths exceeding $E_'d' = \SI{1}{MV\per m}$.
This is a low threshold, in the sense that most electrons detach,
therefore, the effect of increasing it is explored as well.
The movement $\Delta s$ of each electron or anion $i$ is calculated by
\begin{equation}
    \Delta \vec{s}_i = \vec{E}_i \, \mu_i \, \Delta t \,.
    \label{eq:ds}
\end{equation}
The simulation time step
$\Delta t$ is chosen low enough,
typically \SIrange{1}{10}{\pico\second},
to ensure
that $\Delta s$ is less than \SI{0.1}{\um}.
For a positive streamer,
the negative charged species move towards higher field strengths.
Increasing the electric field strength,
increases the kinetic energy an electron gains between colliding with molecules
as well as lowering
the ionization potential (IP) of the molecules~\citep{Smalo2011},
which increases the probability of impact ionization.
As electron attachment processes dominate at low field strengths,
an electric field exceeding $E_'a' = \SI{0.2}{GV/m}$
is required for electron multiplication to be observed in cyclohexane~\citep{Haidara1991}.
The electric field at a streamer head must not only exceed $E_'a'$,
but also be strong enough to cause electron multiplication over a sufficient distance,
for the streamer to propagate.

An electron avalanche occurs
when electron multiplication is dominant
and
the number of electrons $N_'e'$ grows rapidly.
The growth of such an avalanche is modeled as~\citep{Pedersen1989}
\begin{equation}
    \d N_'e' = N_'e' \, \alpha \, \d s \,,
    \label{eq:dN_e}
\end{equation}
where
$\alpha$ is the average number of electrons generated per unit length.
For discharges in gases,
$\alpha$ is assumed to be dependent on
the type of molecules, the density,
and the electric field strength~\citep{Heylen2000}.
Assuming that the same holds for a liquid,
considering a constant liquid density~\citep{Atrazhev1991,Haidara1991},
yields
\begin{equation}
    \alpha = \alpha_'m' \exp \left( - \frac{E_\alpha}{E} \right) \,.
    \label{eq:alpha}
\end{equation}
The maximum avalanche growth $\alpha_'m'$
and the inelastic scattering constant $E_\alpha$
are dependent on the liquid
and are found from experimental data~\citep{Haidara1991,Naidis2015tdei}.
\Cref{eq:dN_e} leads to an exponential growth of electrons in an avalanche,
\begin{equation}
    N_'e' = N_0 \exp \left ( {\int \alpha \, \d s} \right )
          = N_0 \exp Q_'e' \,,
    \label{eq:N_e}
\end{equation}
where $N_0$ is the initial number of electrons,
and $Q_'e'$ is introduced as a measure of the avalanche size.
At each simulation step,
$Q_'e'$ for each avalanche is increased by
\begin{equation}
    \Delta Q = \alpha \, \Delta s
             =  \alpha \, E \, \mu \, \Delta t \,.
    \label{eq:DQ}
\end{equation}
For discharges in gases it is assumed that
an electron avalanche becomes unstable when the electron number $N_'e'$
exceeds some threshold $N_'c'$,
which is known as
the Townsend--Meek avalanche-to-streamer criterion~\citep{Pedersen1989}.
In the model, an avalanche obtaining this criterion is removed
and its position is considered as a part of the streamer channel.
Assuming that an avalanche starts from a single electron,
the criterion $N_'e' > N_'c'$ is rewritten as
\begin{equation}
    Q_'e' = \ln N_'e' > Q_'c' \,.
    \label{eq:Q_e}
\end{equation}
The Meek constant $Q_'c'$ is typically 18 in gases~\citep{Pedersen1989,Lowke2003},
but the value is expected to be higher in liquids
since the denser media has a higher breakdown strength,
and creation of higher electric fields requires more electrons.
However,
a recent study on liquids
found values in the range 5 to 20
when evaluating a number of experiments~\citep{Naidis2015tdei}.
Another study found $Q_'c' = 23$,
or an avalanche size of about $10^{10}$ electrons,
by considering the field required for propagation~\citep{Ingebrigtsen2009},
in contrast to the field required for initiation,
which is more common.

%


\subsection{Additives}\label{sec:additives}

Additives with low IP
have proven to facilitate the propagation of 2nd mode streamers,
since such additives lower the voltage required
for propagation and for breakdown,
whereas they increase the voltage required
for 4th mode streamers~\citep{Lesaint2016}.
This is likely a consequence of an increased number of branches,
which may increase the electrostatic shielding
and thereby reducing the electric field
at the streamer heads~\citep{Lundgaard1998,Lesaint2000}.
To account for the effect of low-IP additives on electron avalanche growth,
the mole fraction $c_'n'$ of the additive
and the IP difference between the base liquid $I_'b'$ and the additive $I_'a'$,
is used to modify the expression for $\alpha$ in \cref{eq:alpha} as~\citep{Ingebrigtsen2009}
\begin{equation}
    \alpha' =
        \alpha \, \Big(
            1 - c_'n' + c_'n' \, e^{k_\alpha (I_'b' - I_'a') } \Big) \,,
    \label{eq:alpha'}
\end{equation}
where the parameter $k_\alpha = \SI{2.8}{eV^{-1}}$ is estimated from experiments%
\citep{Ingebrigtsen2009}.
For example,
an additive with an IP difference of \SI{3.1}{eV} from the base liquid,
in a concentration $c_'n'$ of \SI{0.1}{\%},
yields $\alpha' = 6.9 \alpha$.
\Cref{eq:alpha'} is derived assuming that
ionization is caused by electrons in the exponentially decaying,
high-energy tail of a Maxwellian distribution,
and that the introduction of an additive
does not significantly change the energy distribution~\citep{Ingebrigtsen2009}.

%


\subsection{Streamer representation}

The model focuses on the processes occurring in front of the streamer.
The streamer is represented by a collection of hyperboloids,
approximating the electric field in front of the streamer.
The streamer channel, and in particular its dynamics,
is not included in the model.
The streamer hyperboloids are referred to as ``streamer heads'',
and the initial streamer consists of only one streamer head: the needle.
The needle, one other streamer head, and relevant variables,
are shown in \cref{fig:geometry}.

The potential $V$ at position $\vec{r}$
is given by a superposition
of the potential $V_i$ in \cref{eq:ps_Vi}
of each streamer head,
\begin{equation}
    V (\vec{r}) = \sum_i k_i \, V_i (\vec{r}) \,,
    \label{eq:V_r}
\end{equation}
where the coefficients $k_i$
are introduced to account for electrostatic shielding between the heads.
The electric field is found in a similar manner,
\begin{equation}
    \vec{E} (\vec{r}) = \sum_i k_i \, \vec{E}_i (\vec{r}) \,,
    \label{eq:E_r}
\end{equation}
where $\vec{E}_i$ in \cref{eq:ps_Ei} is the electric field
arising from streamer head $i$.
The electric field arising from a streamer head
is strongly dependent on its tip radius $r_'p'$.
Experiments have shown that there exists a critical tip radius
for the inception of 2nd mode streamers,
which is $r_'p' = \SI{6}{\micro\metre}$ for cyclohexane%
~\citep{Gournay1993,Yamashita1998}.

When an electron avalanche meets the Townsend--\allowbreak{}Meek criterion
in~\cref{eq:Q_e},
a new streamer head is added at the position of the avalanche.
The potential at the tip of streamer head $i$ is given by
\begin{equation}
    V_i(\vec{r}_i) = V_0 - E_'s' \, \ell_i \,,
    \label{eq:Vi_ri}
\end{equation}
where
$V_0$ is the potential at the needle,
$E_'s'$ is the electric field within the streamer channel,
and $\ell_i$ is the distance
from the tip of the needle to the tip of streamer head $i$,
\begin{equation}
    \ell_i = \abs{ \vec{r}_i - d_'g' \, \hat{z}} \,,
\end{equation}
again
see \cref{fig:geometry} for definitions.
\Cref{eq:Vi_ri} is used to find $C_i$ through \cref{eq:ps_Ci}.

The shielding coefficients $k_i$ ensure that
the combined potential of all the streamer heads equals
the potential at the tip of each streamer head,
\begin{equation}
    V (\vec{r}_i)
        = \sum_j k_j \, V_j (\vec{r}_i)
        \approx V_i (\vec{r}_i)
    \,,
    \label{eq:V_i_opt1}
\end{equation}
and are obtained by a non-negative least squares (\textsc{nnls}) routine~\citep{Lawson1995}.
The problem actually solved numerically
is stated in a slightly different form.
Defining
\begin{equation}
    M_{ij}
        = \frac{V_j (\vec{r}_i)}{V_j (\vec{r}_j)}
        = \frac{\ln \tan \left( \frac{1}{2} \nu_j \! \left( \vec{r}_i \right) \right)}
               {\ln \tan \left( \frac{1}{2} \nu_j \! \left( \vec{r}_j \right) \right)} \,,
        \label{eq:M_V}
\end{equation}
which only depend on the geometry and not on the potentials,
\cref{eq:V_i_opt1} is rewritten as
\begin{equation}
    V_i (\vec{r}_i)
        \approx \sum_j M_{ij} \, k_j \, V_j (\vec{r}_j) \,,
        \label{eq:V_i_opt2}
\end{equation}
which is computationally more convenient to solve.

\begin{figure}
    \centering
    \includegraphics[width=1.0\linewidth]{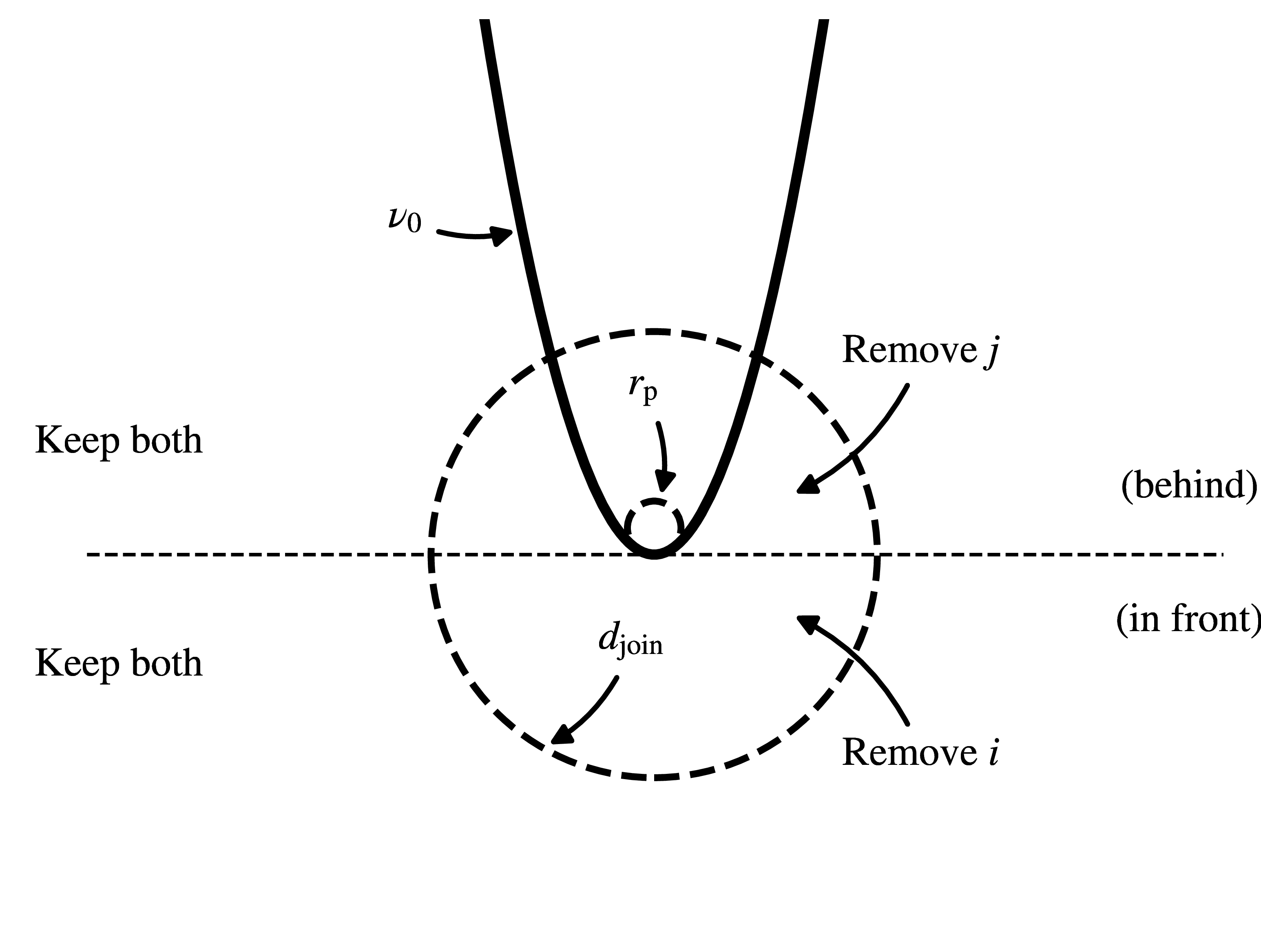}
    \caption{
        For given a streamer head $i$ (shown),
        other positions are considered to be
        within, behind, in front, and/or within join distance.
    }
    \label{fig:trim_tip}
\end{figure}

It is desirable to keep the number of streamer heads to a minimum
since it is expensive to calculate the electric field from a head.
Also optimization of the potential becomes more difficult and unstable
as it tend to become a more overdetermined problem with more heads present,
especially when the heads are close or ``within'' each other.
Streamer heads located within another streamer head are removed,
that is,
if
\begin{equation}
    \nu_i(\vec{r}_j) < \nu_i(\vec{r}_i) \,,
    \label{eq:remove_within}
\end{equation}
then
streamer head $j$ is removed,
which is the same as being above the $\nu_0$-line in \cref{fig:trim_tip}.
In addition, if the tip of one streamer head
is within a certain distance $d_'m'$
of the tip of another streamer head,
\begin{equation}
    \abs{\vec{r}_i - \vec{r}_j} < d_'m' \,,
    \label{eq:remove_close}
\end{equation}
the heads are merged and
only the streamer head closest to the plane is kept
(see \cref{fig:trim_tip}).
Physically, this is motivated as charge transferred
from one streamer head
to another located closer to the grounded plane.
Finally,
since fewer heads implies less calculation and faster simulations,
streamer heads with a shielding coefficient below a given threshold,
\begin{equation}
    k_i < k_'c' \,,
    \label{eq:kc}
\end{equation}
are also removed.
When $k_'c'$ is chosen sufficiently low,
only streamer heads that are
to a large degree
shielded by other heads are removed,
and removing them have thus little effect on the simulation results.

The streamer consist of one or more heads as it propagates.
When a new head is added,
the conditions \cref{eq:remove_within,eq:remove_close}
are used to evaluate whether the new heads should be kept
and whether any of the existing heads should be removed.
A new head added at a sufficient distance from the existing head(s)
can initiate streamer branching.
However, for the actual branching to occur,
the streamer must be able to propagate (add new heads)
both from the new head and from the existing head(s).
The result is then that the streamer at some point
grows in two directions at the same time.
This occurs rarely,
since the leading streamer head
shields the potential of the other heads
and reduces the probability of propagation from those heads.

%


\subsection{Region of interest}

\begin{figure}[t!]
    \centering
    \includegraphics[width=1.0\linewidth]{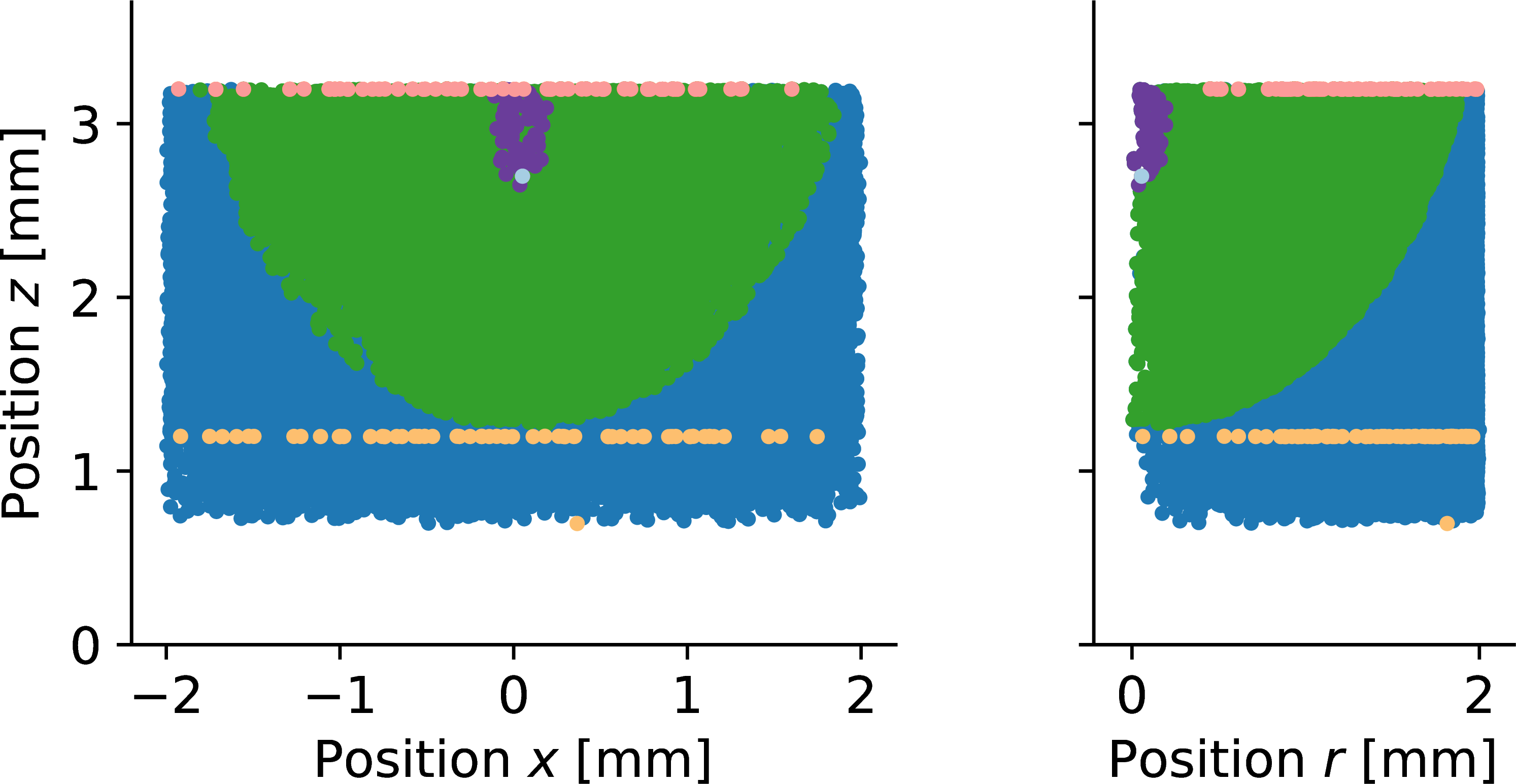}
    \caption{
        Region of Interest,
        $xz$- and $rz$-projection.
        Each seed is represented by a dot;
        anion (blue),
        electrons (green),
        avalanches (purple),
        behind ROI (pink),
        newly placed (tan),
        and a single critical
        (light blue).
    }
    \label{fig:seed_distribution}
\end{figure}

\begin{figure}[t!]
    \centering
    \includegraphics[width=1.0\linewidth]{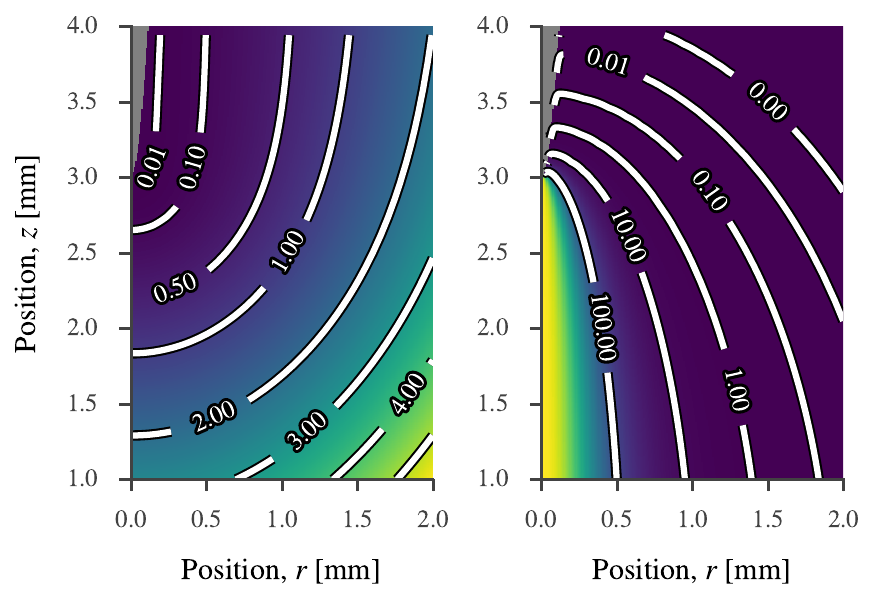}
    \caption{
        Time to collision $t_'i'$ (left, in milliseconds)
        and maximum avalanche size $Q_'i'$ (right),
        for an electron originating at a given position.
        The needle hyperbola is shown in gray.
        For a gap distance of \SI{3}{\mm},
        a tip radius of \SI{6}{\micro\metre},
        and at a potential of \SI{100}{\kV}.
    }
    \label{fig:ti_qi}
\end{figure}

Anions, electrons, and avalanches
are here referred to as ``seeds''.
The seeds are placed as anions,
but can become electrons or avalanches,
depending on the local electric field strength,
which is illustrated in \cref{fig:seed_distribution}.
To save computational cost,
especially for simulations in large gaps,
seeds are limited to a region of interest (ROI)
surrounding the leading tip, see \cref{fig:seed_distribution}.
The ROI is a cylinder defined by a radius from the centerline
($x^2 + y^2 = r^2$),
a distance in front of the leading streamer head,
and a distance behind the leading head.
Seed avalanches that obtain a critical size,
seeds that collide with a streamer head,
and seeds that fall behind the ROI,
are removed
and replaced by new seeds.
A new seed is placed one ROI length from the old seed
in the $z$-direction,
with random placement within the ROI radius
for the $x$- and $y$-coordinates.
The seed density is thus kept constant
as the ROI moves together with the leading streamer head.

Removing or rearranging the seeds does not change the electric field,
since the charge from the seeds is not included in the Laplacian field.
Charge from single cations, anions, or electrons
should not have a big influence,
but the charge from electrons and cations
created by electrons avalanches
is also ignored,
and this is a major simplification.
An avalanche colliding with the streamer
is shielded by the streamer and
does not contribute to the streamer propagation.
A critical avalanche,
however,
propagates the streamer potential to its position.
In any case, when an avalanche is removed,
its charge is considered as absorbed by the streamer.

For a given configuration,
it is possible to calculate the time $t_'i'$
for an electron to travel from
a given point to the needle.
This is achieved by numeric integration of $v_'d'^{-1} \d l$
along an electric field line (constant $\mu$),
using $h_\nu = \d l / \d \nu$ (cf.~\cref{eq:h_nu}),
\begin{equation}
    t_'i' = \int\limits_'position'^'needle' \frac{h_\nu}{v_'d'} \, \d \nu \,.
    \label{eq:ti}
\end{equation}
Similarly, the maximum avalanche size $Q_'i'$, is computed by
\begin{equation}
    Q_'i' = \int\limits_'position'^'needle' \alpha \, h_\nu \, \d \nu \,.
    \label{eq:Qi}
\end{equation}
An illustration of \cref{eq:ti,eq:Qi} is found in \cref{fig:ti_qi}.
Both $v_'d'$ in \cref{eq:v_d} and $\alpha$ in \cref{eq:alpha}
are functions of the electric field $E$ in~\cref{eq:ps_Ei},
which makes numeric integration straightforward in prolate spheroid coordinates.
The time, $t_'i'$, provides an indication of how large the ROI should be.
Given that a slow streamer may propagate at
\SI{1}{\km \per \s} = \SI{1}{\mm \per \us},
the ROI should be chosen so wide
that seeds on the sides does not have enough time to collide
with the passing streamer.
According to \cref{fig:ti_qi},
a width of \SI{1.5}{\mm} gives about \SI{1}{\us} before collision,
both from the sides and from below.
As the streamer should propagate about the same length,
or more,
in this time,
is a reasonable value.
However, a somewhat wider ROI should be used to account for
a streamer propagating off-center, and for branched propagation.
Further, \cref{fig:ti_qi} shows that $Q_'i'$ is large in the front,
but quickly declines for seeds behind the streamer head.
This gives an indication on how far behind the streamer head
an avalanche may obtain critical size,
which is how far behind the streamer head
it is interesting to extend the ROI.
However, the ROI should also extend
far enough behind the leading streamer head
to enable the propagation of secondary branches.
Even though $t_'i'$ and $Q_'i'$ give good indications
of how big the ROI should be,
it is important to verify the settings after the simulation,
or vary the ROI to verify that the results are not affected.

%


\subsection{Parameters}

\begin{table}[!t]
    \caption{Model parameters, physical.}
{%
\centering
\renewcommand{\arraystretch}{1.1}%
\begin{tabularx}{1.0\linewidth}{
    >{\raggedright}X
    >{\centering\arraybackslash\hsize=.15\hsize}X
    >{\centering\arraybackslash\hsize=.40\hsize}X
    }%
    \toprule
        Gap distance
    &   $d_'g'$
    &   \SI{3.0}{\milli\metre}
    \\
        Applied voltage (varies)
    &   $V_'n'$
    &   $-$
    \\
        Needle tip curvature
    &   $r_'n'$
    &   \SI{6.0}{\micro\metre}
    \\
        Streamer tip curvature~\citep{Gournay1993}
    &   $r_'s'$
    &   \SI{6.0}{\micro\metre}
    \\
        Field in streamer~\citep{Lesaint1988,Massala1998}
    &   $E_'s'$
    &   \SI{2.0}{\kilo\volt\per\milli\metre}
    \\
        Electron detachment threshold
    &   $E_'d'$
    &   \SI{1.0}{\mega\volt\per\metre}
    \\
        Avalanche threshold~\citep{Haidara1991}
    &   $E_'a'$
    &   \SI{0.2}{\giga\volt\per\metre}
    \\
        Scattering constant~\citep{Haidara1991}
    &   $E_\alpha$
    &   \SI{3.0}{\giga\volt\per\metre}
    \\
        Max avalanche growth~\citep{Haidara1991}
    &   $\alpha_'m'$
    &   \SI[per-mode=reciprocal]{200}{\per\micro\metre}
    \\
        Meek constant~\citep{Ingebrigtsen2009}
    &   $Q_'c'$
    &   \SI{23}{ }
    \\
        Electron mobility~\citep{Allen1976,Schmidt1977}
    &   $\mu_'e'$
    &   \SI{45}{\milli\metre^{2}\per{\volt\second}}
    \\
        Anion mobility~\citep{Dumitrescu2001}
    &   $\mu_'ion'$
    &   \SI{0.30}{\milli\metre^{2}\per{\volt\second}}
    \\
        Ion conductivity~\citep{Alj1985}
    &   $\sigma_'ion'$
    &   \SI{0.20}{\pico\siemens\per\metre}
    \\
        Base liquid IP~\citep{Davari2014}
    &   $I_'b'$
    &   \SI{10.2}{\eV}
    \\
        Additive IP~\citep{Davari2013}
    &   $I_'a'$
    &   \SI{7.1}{\eV}
    \\
        Additive IP diff. factor~\citep{Ingebrigtsen2009}
    &   $k_\alpha$
    &   \SI{2.8}{\eV^{-1}}
    \\
        Additive number density
    &   $c_'a,n'$
    &   \SI{0.0}{}
    \\
    \bottomrule%
\end{tabularx}%
}%

    \label{tab:tab_param_phys}
\end{table}

\begin{table}[!t]
    \caption{Model parameters, algorithm.}
{%
\centering
\renewcommand{\arraystretch}{1.1}%
\begin{tabularx}{1.0\linewidth}{
    >{\raggedright}X
    >{\centering\arraybackslash\hsize=.15\hsize}X
    >{\centering\arraybackslash\hsize=.40\hsize}X
    }%
    \toprule
        Streamer head merge distance
    &   $d_'m'$
    &   \SI{50}{\micro\metre}
    \\
        Potential shielding threshold
    &   $k_'c'$
    &   \SI{0.10}{}
    \\
        Time step
    &   $\Delta t$
    &   \SI{1.0}{\pico\second}
    \\
        Micro step number
    &   $N_"msn"$
    &   \SI{100}{}
    \\
        ROI -- behind leading head
    &   $z^{+}_"roi"$
    &   \SI{0.5}{\milli\metre}
    \\
        ROI -- in front of leading head
    &   $z^{-}_"roi"$
    &   \SI{1.5}{\milli\metre}
    \\
        ROI -- radius from center
    &   $r_"roi"$
    &   \SI{2.0}{\milli\metre}
    \\
        Stop -- low streamer speed
    &   $v_'min'$
    &   \SI{100}{\metre\per\second}
    \\
        Stop -- streamer close to plane
    &   $z_'min'$
    &   \SI{50}{\micro\metre}
    \\
        Stop -- avalanche time
    &   $t^'ava'_'max'$
    &   \SI{100}{\nano\second}
    \\
    \bottomrule%
\end{tabularx}%
}%

    \label{tab:tab_param_alg}
\end{table}

The model parameters may be divided in two groups:
physical parameters and parameters for the numerical algorithm.
The values of the physical parameters
summarized in \cref{tab:tab_param_phys}
are given by the properties of the simulated experiment or
based on values available in the literature for
the base liquid (cyclohexane)
and the additive (dimethylaniline).
Since not all the parameter values are available and some are uncertain,
a sensitivity analysis is carried out in this work to investigate
the influence of individual parameters.
%
Parameter values needed by the simulation algorithm,
which are not based on physical properties,
are given in \cref{tab:tab_param_alg}
and include the size of the ROI
and certain criteria for stopping a simulation.

The initial setup is given by
$V_'n'$,
$d_'g'$, and
$r_'n'$.
Then the number fraction of seeds $n_'ion'$
is calculated using
$\mu_'ion'$ and $\sigma_'ion'$,
according to \cref{eq:n_ion},
and whether a seed is considered as
an anion, an electron, or an avalanche
is given by $E_'d'$ and $E_'a'$.
The electron multiplication probability is given by \cref{eq:alpha},
using $E_\alpha$ and $\alpha_'m'$.
If an additive is present,
then \cref{eq:alpha'} is also applied,
where
$I_'b'$,
$I_'a'$,
$c_'a,n'$, and
$k_\alpha$
are used.
\Cref{eq:DQ} gives the growth of an avalanche,
using $\Delta t$ and $\mu_'e'$.
Finally, the Townsend--Meek criterion,
stated in \cref{eq:Q_e},
uses $Q_'c'$
to evaluate whether the avalanche has obtained a critical size.
The streamer branching is regulated by
$d_'m'$ and
$k_'c'$, by \cref{eq:remove_close,eq:kc},
while the streamer head potential,
and thus also the electric field at the tip,
is dependent on
$E_'s'$ and
$r_'s'$
through \cref{eq:Vi_ri}.

%


\subsection{Algorithm}\label{sec:model_algorithm}{

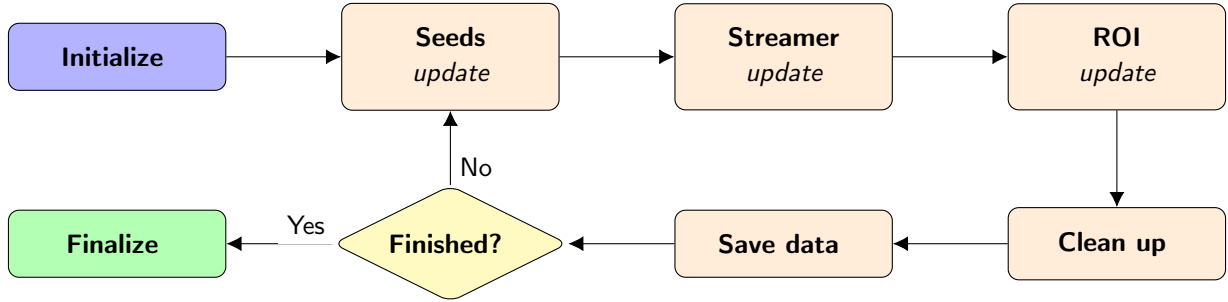
\begin{figure*}[!t]
  \centering

\newcommand{\headts}{\bfseries}  
\newcommand{\smallts}{\itshape}  

\begin{tikzpicture}[
    node distance=2 cm,
    every node/.style={fill=white, font=\sffamily},
    align=center,
    ]

  \node (read)  [activityStarts]{
    \headts  Initialize
    };

  \node (seed)  [process, right of=read,  xshift=15 ex]{
    \headts  Seeds\\
    \smallts update
    };

  \node (strm) [process, right of=seed, xshift=15 ex]{
    \headts  Streamer\\
    \smallts update
    };
  \node (roi)[process, right of=strm, xshift=15 ex]{
    \headts  ROI\\
    \smallts update
    };

  \node (clen)[process, below of=roi, yshift=-1 ex]{
    \headts  Clean up
    };

  \node (save)[process, left of=clen, xshift=-15 ex]{
    \headts  Save data
    };

  \node (brek)[decision, left of=save, xshift=-15 ex]{
    \headts  Finished?
    };

  \node (done) [activityRuns, left of=brek, xshift=-15 ex]{
    \headts  Finalize
    };

  \draw[->]  (read) -- (seed);
  \draw[->]  (seed) -- (strm);
  \draw[->]  (strm) -- (roi);
  \draw[->]  (roi) -- (clen);
  \draw[->]  (clen) -- (save);
  \draw[->]  (save) -- (brek);
  \draw[->]  (brek) -- node [above, near start]{Yes} (done);
  \draw[->]  (brek) -- node [right, near start]{No} (seed);

\end{tikzpicture}
  \caption{
      The main steps of the simulation algorithm.
      The algorithm for the seeds is detailed in \cref{fig:alg_seeds}.
      See \cref{sec:model_algorithm} for further details on each step.
      }
    \label{fig:alg_main}
\end{figure*}

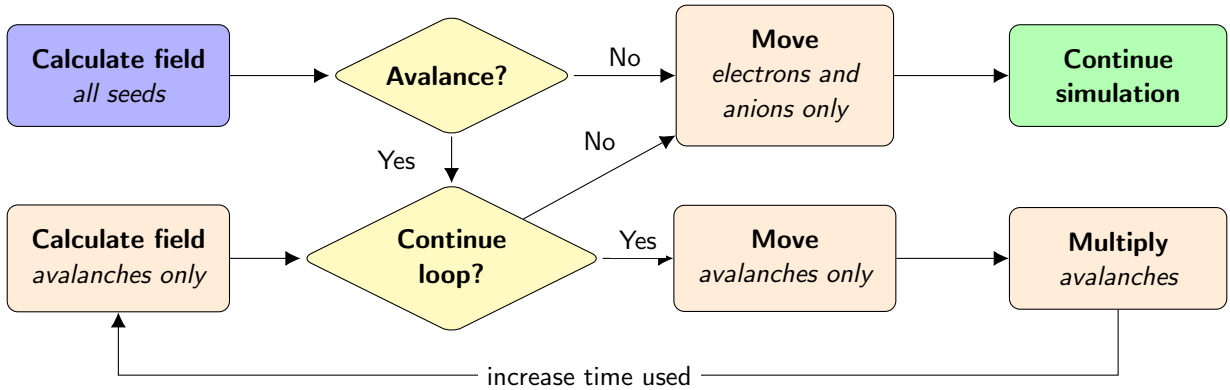
\begin{figure*}[!t]
  \centering

\newcommand{\headts}{\bfseries}  
\newcommand{\smallts}{\itshape}  

\begin{tikzpicture}[
    node distance=2 cm,
    every node/.style={fill=white, font=\sffamily},
    align=center,
    ]

  \node (1a)  [activityStarts]{
    \headts  Calculate field\\
    \smallts all seeds
    };
  \node (2a)  [decision, right of=1a,  xshift=15 ex]{
    \headts  Avalance?
    };
  \node (3a) [process, right of=2a, xshift=15 ex]{
    \headts  Move\\
    \smallts electrons and\\
    \smallts anions only
    };
  \node (4a)[activityRuns, right of=3a, xshift=15 ex]{
    \headts  Continue\\
    \headts  simulation
    };
  \node (1b)  [process, below of=1a, yshift=-12]{
    \headts  Calculate field\\
    \smallts avalanches only
    };
  \node (2b)  [decision, right of=1b,  xshift=15 ex]{
    \headts  Continue\\
    \headts  loop?
    };
  \node (3b) [process, right of=2b, xshift=15 ex]{
    \headts  Move\\
    \smallts avalanches only
    };
  \node (4b)[process, right of=3b, xshift=15 ex]{
    \headts  Multiply\\
    \smallts avalanches
    };

  \draw[->]  (1a) -- (2a);
  \draw[->]  (2a) -- node [above]{No} (3a);
  \draw[->]  (3a) -- (4a);
  \draw[->]  (1b) -- (2b);
  \draw[->]  (2b) -- node [above]{Yes} (3b);
  \draw[->]  (3b) -- (4b);

  \draw[->]  (2a) -- node [left=1 em]{Yes} (2b);
  \draw[->]  (2b) -- node [above=1.5 ex]{No} (3a);
  \draw[->]  (4b) |- (39 ex, -30 ex) node []{increase time used} -| (1b);

\end{tikzpicture}
  \caption{
      Algorithm for moving and multiplying seeds.
      This is the block labeled ``Seeds'' in \cref{fig:alg_main}.
      See \cref{sec:model_algorithm} for details on each step.
      }
    \label{fig:alg_seeds}
\end{figure*}

A simulation begins by reading an input file
that is used to initialize the various data classes
used by the program,
including random placement of seeds within the ROI,
thereafter, a loop is executed until the simulation is complete.
These main steps are shown in \cref{fig:alg_main}.
The first and most expensive step of the algorithm
is the update of the seeds,
which is detailed in \cref{fig:alg_seeds}.
First,
the electric field is calculated for all seeds
(each anion, electron, and avalanche).
All the avalanches are treated separately in a loop,
where they are moved,
the electrons are multiplied,
and the field is calculated for their new positions.
This loop, in  \cref{fig:alg_seeds},
is performed until either
$N_"msn"$ steps are done,
an avalanche becomes critical
(obtaining the Townsend--Meek criterion),
or an avalanche collides with the streamer.
Then,
all other seeds (anions and electrons) are moved,
using a time step
equal to the total time used by the avalanches.
The next step
in \cref{fig:alg_main}
is to update the streamer structure.
Any critical avalanches are added to the streamer,
and the streamer structure is optimized
by removing heads using \cref{eq:remove_within,eq:remove_close}
and correcting the scaling using \cref{eq:V_i_opt2}
to set $k_i$ for each streamer head.
Thereafter,
if there is a new leading streamer head,
the ROI is updated.
In the ``clean-up'' part,
seeds behind the ROI,
critical seeds,
and seeds that have collided with the streamer,
are removed and replaced by new seeds.
A number of criteria can be set
to determine when the simulation loop
in \cref{fig:alg_main}
should end.
For instance,
total simulation time,
total CPU time,
and number of iterations.
However,
simulations presented in this work ended for one of three reasons:
the leading head reached the planar electrode ($z_i < z_'min'$),
low propagation speed ($ < v_'min'$),
or long time between critical avalanches ($ > t^'ava'_'max'$).
The final step of the loop is saving data,
and
finalizing a simulation ensures that
all temporary data is properly saved to files.

The implementation has been done in Python~\citep{Oliphant2007}
using NumPy~\citep{VanDerWalt2011} extensively.
During initialization,
the seed for random numbers is set in NumPy
to ensure reproducible results.
The input parameters are given in a JSON-formatted file,
which is used for initiation of the simulation.
Simulation results are
saved with Pickle and
illustrated using Matplotlib~\citep{Hunter2007}.

}

}

%


\section{Simulation results and discussion}\label{sec:results}{

The model involves numerous parameters,
some of which is given by the experimental setup (e.g. gap distance),
others by properties of the liquids (e.g. mobilities),
and some are purely for the simulation procedure (e.g. time step).
In the first part,
the default parameters
given by \cref{tab:tab_param_phys,tab:tab_param_alg}
show the basic behavior of the model.
Thereafter,
a sensitivity analysis is presented,
indicating the influence of various parameters.
Mainly the propagation speed is used to indicate the differences,
but
the number of streamer heads,
their scaling $k_i$,
the propagation length,
and
the degree of branching
are also investigated.
Ten simulations are carried out at each voltage,
using the numbers 1 to 10 in the random number generator
generating different initial configurations
of the seed distribution.


\subsection{Simulation baseline}\label{sec:results_baseline}

\begin{figure}[t!]
    \centering
    \includegraphics[width=0.95\linewidth]{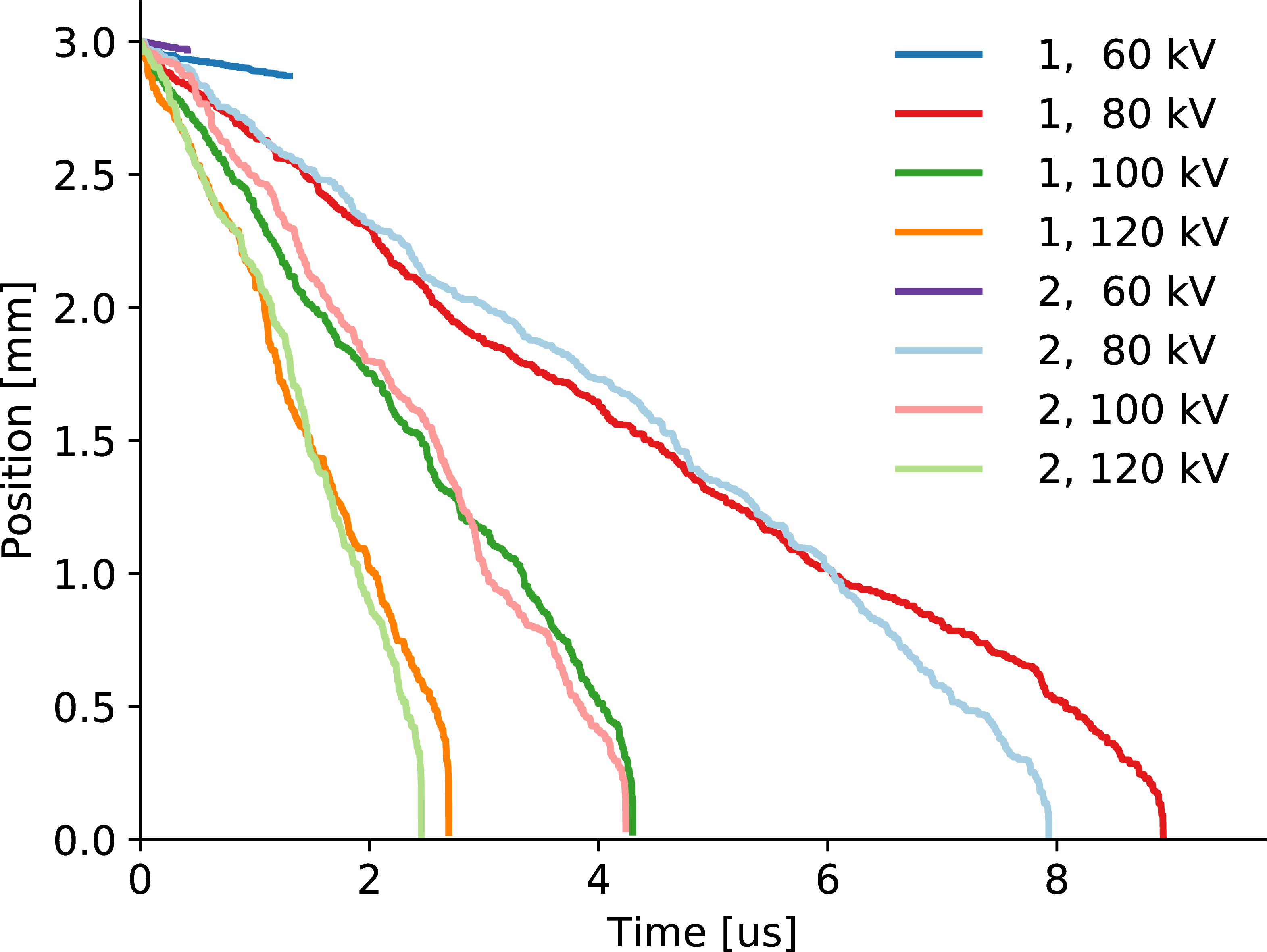}
    \caption{
        Streak plots, time spent versus leading head position,
        for two simulations (different initial random numbers) at each voltage.
        The streamers start a the position of the needle,
        $z = d_'g' = \SI{3.0}{\mm}$.
        }
    \label{fig:fig_kc_multi_streak_zapsim_930_stat}
\end{figure}

\begin{figure}[t!]
    \centering
    \includegraphics[width=0.95\linewidth]{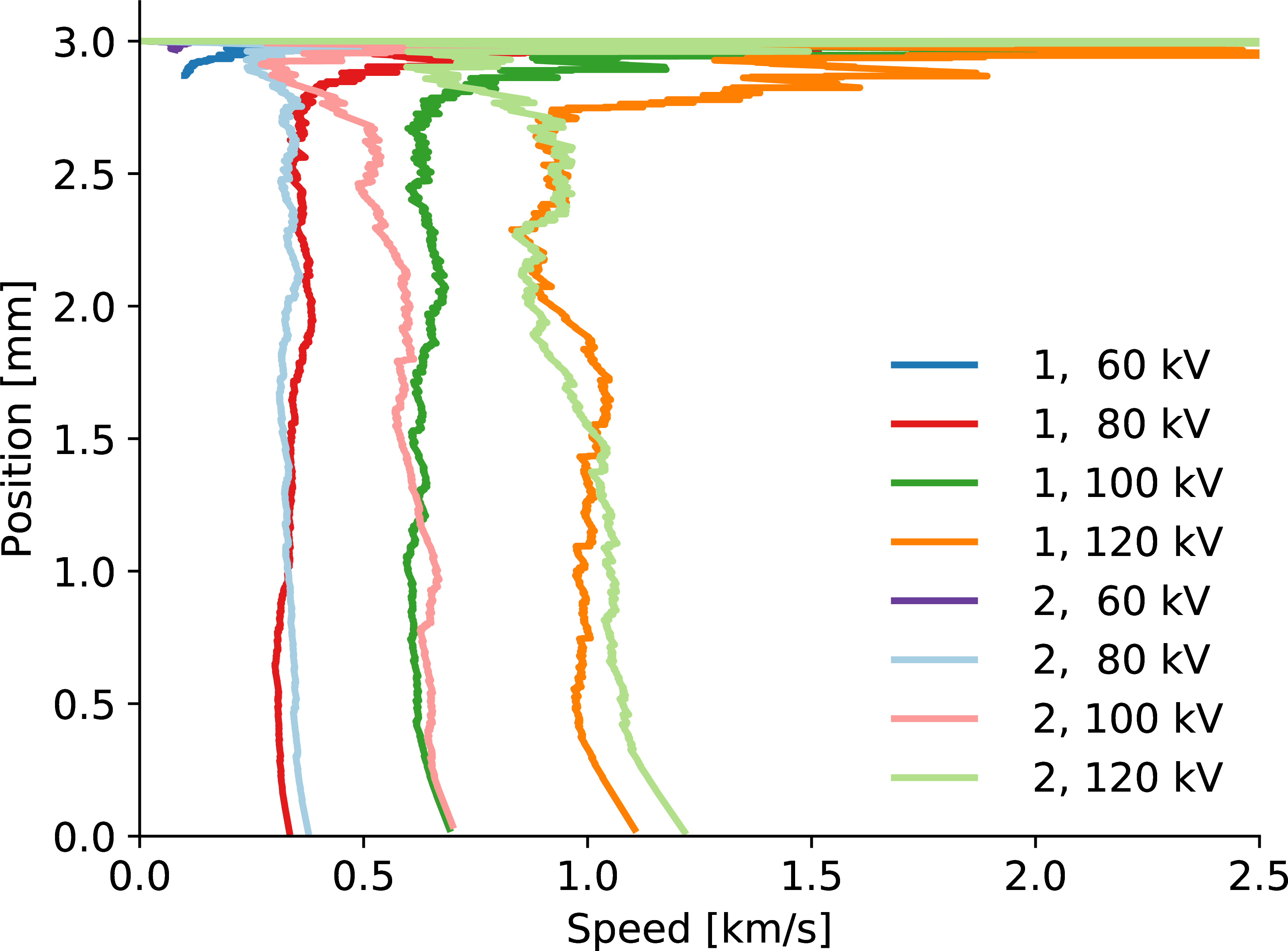}
    \caption{
        Streamer average speed versus leading head position,
        that is,
        the average gradient of the ``streaks'' shown in
        \cref{fig:fig_kc_multi_streak_zapsim_930_stat}.
        }
    \label{fig:fig_kc_multi_speed_avg_zapsim_930_stat}
\end{figure}

\begin{figure}[t!]
    \centering
    \includegraphics[width=0.95\linewidth]{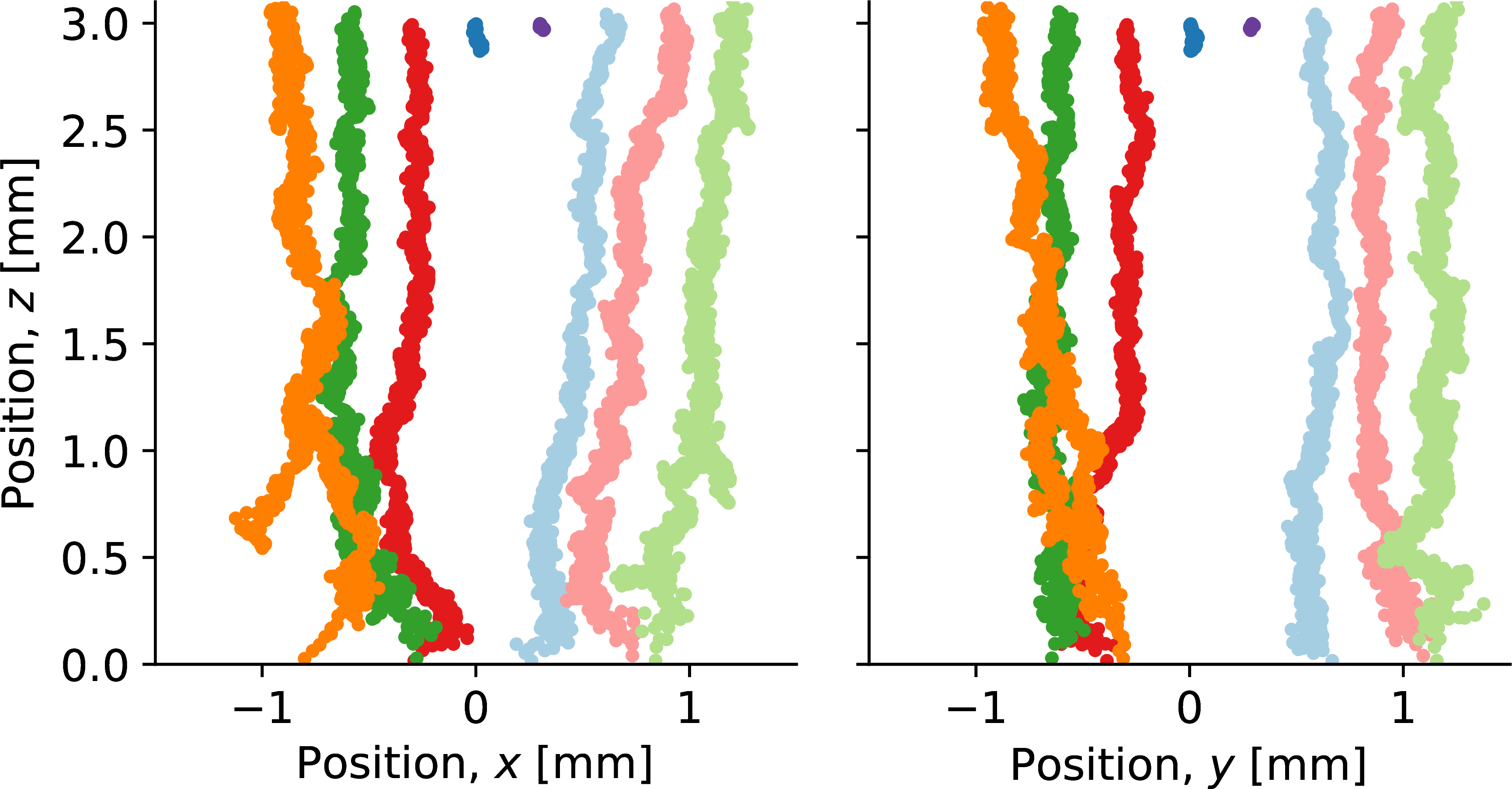}
    \caption{
        Streamer trails,
        $xz$- and $yz$-projection for a range of voltages
        \SIrange{60}{120}{\kV},
        using the same legend as in \cref{fig:fig_kc_multi_streak_zapsim_930_stat}.
        Each dot represents the position of a streamer head
        at some point of the propagation.
        The streamers are plotted with an offset
        to improve the readability.
        }
    \label{fig:fig_kc_multi_trail_zapsim_930_stat}
\end{figure}

\begin{figure}[t!]
    \centering
    \includegraphics[width=0.95\linewidth]{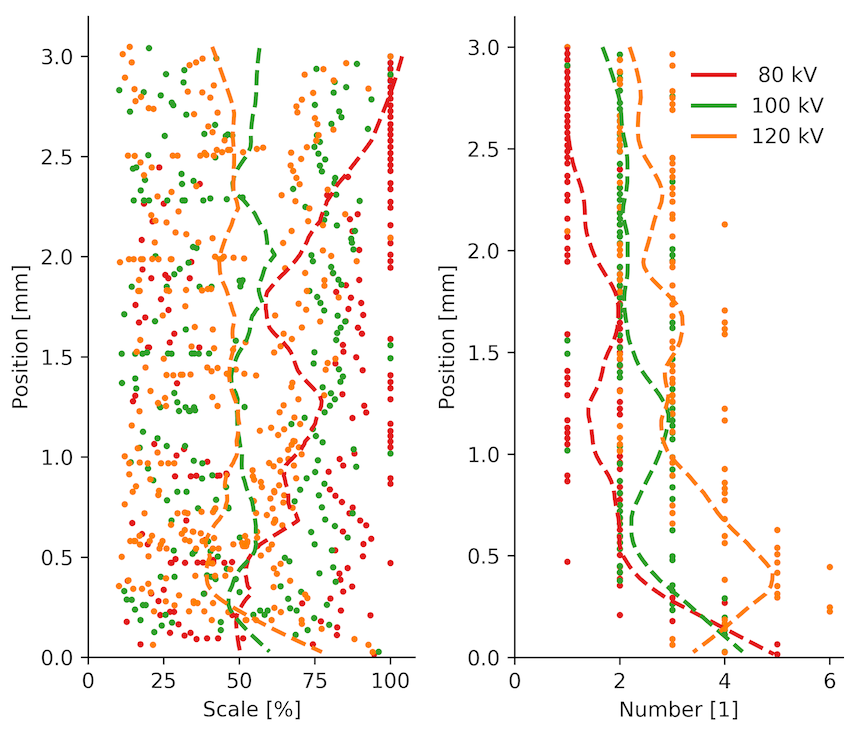}
    \caption{
        Actual streamer head scale $k_i$ (left)
        and total number of streamer heads (right).
        Data are taken every 1~\% of the gap.
        The dashed lines are moving averages
        calculated by loess-regression~\citep{Cleveland1979}.
        }
    \label{fig:fig_kc_multi_tipscale_tipsno_zapsim_930_gp1}
\end{figure}

\begin{figure}[t!]
    \centering
    \includegraphics[width=0.95\linewidth]{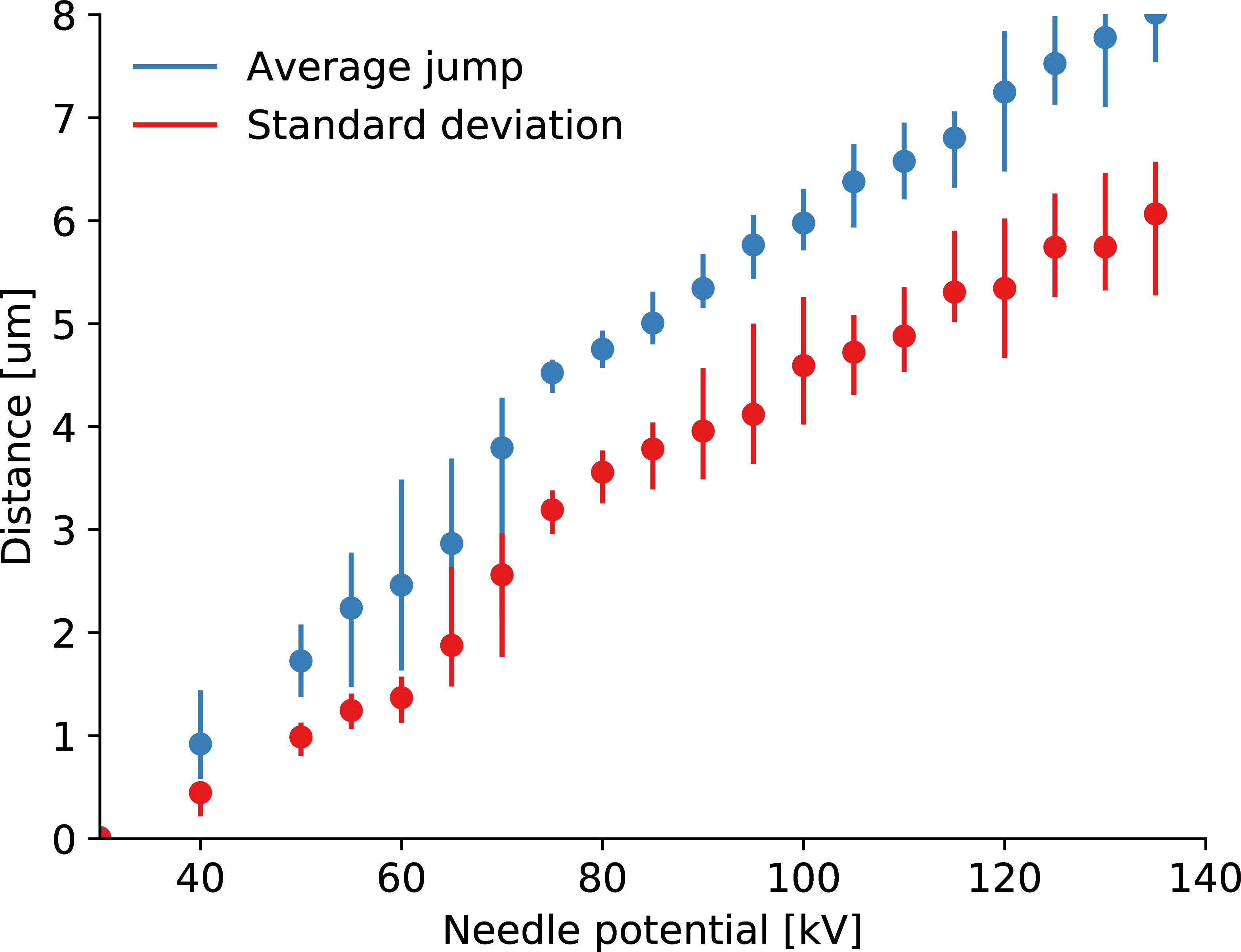}
    \caption{
        The leading streamer head is moved in a sequence of discrete ``jumps''
        in the $z$-direction.
        The average jump length and
        the standard deviation of the jumps
        are found for each individual simulation.
        The dotted lines are interpolated to the average,
        and the bars covers the minimum and maximum values
        for ten simulations at the same voltage.
        }
    \label{fig:fig_kc_zapsim_stat_v_jda_jds_sst(c)}
\end{figure}

Simulations have been performed
for a range of voltages,
using the parameters in \cref{tab:tab_param_phys,tab:tab_param_alg}.
These simulations are used as a baseline
in the sensitivity analysis.
As seen from the streak plots in \cref{fig:fig_kc_multi_streak_zapsim_930_stat},
a voltage exceeding \SI{60}{\kV} is needed for a breakdown.
For lower voltages,
the streamer propagates less than \SI{100}{\micro\meter}
before the simulation is terminated,
either because of waiting too long for an avalanche
or because of very slow propagation speed.
Above the breakdown voltage,
the time to breakdown is reduced as the voltage is increased,
and the streamers tend to
accelerate towards the end of their propagation.
The average propagation speed,
shown in \cref{fig:fig_kc_multi_speed_avg_zapsim_930_stat}
tells a similar story,
but it also indicates that the propagation speed
slows down a bit after the first few steps.
The speed reduction is possibly due to branching,
however, by looking at the streamer in
\cref{fig:fig_kc_multi_trail_zapsim_930_stat},
it is clear that the degree of branching is very low,
but the streamer gets thicker with increasing voltage.
This implies that
even though branching is not apparent,
there are several streamer heads present.
The number of streamer heads may increase
when the electric field strength increases
(at higher voltages or closer to the plane)
as seen in \cref{fig:fig_kc_multi_tipscale_tipsno_zapsim_930_gp1}.
Values of $k_i$ lower than one implies that
the streamer heads shield each other to some degree (cf. \cref{eq:V_r}),
as seen in \cref{fig:fig_kc_multi_tipscale_tipsno_zapsim_930_gp1},
but not enough to stop a propagating streamer.
It is of interest to investigate how the
leading head is affected by shielding,
and
the average scaling indicates this.
The propagation speed can be described by
the time it takes to get a critical avalanche
in front of the leading streamer head
combined with the distance the leading head is moved,
where the latter is presented
in \cref{fig:fig_kc_zapsim_stat_v_jda_jds_sst(c)}.
Increased voltage increases
both the maximum and the average propagation ``jumps'',
especially when the streamer is in the final part of the gap.

The propagation speeds in \cref{fig:fig_kc_multi_speed_avg_zapsim_930_stat}
are somewhat low for 2nd mode streamers,
which should be
\SIrange{1}{10}{\kilo\metre\per\second}~\citep{Lesaint2016}.
Many, if not most, of the simulation parameters
affects the propagation speed.
In the case of the electron mobility $\mu_'e'$,
it is easy to see that
the propagation speed is directly proportional to $\mu_'e'$,
since it only affects the movement of the electrons (cf. \cref{eq:ds}).
For most other parameters, it is not that simple.


\subsection{Effect of avalanche parameters}

\begin{figure}[t!]
    \centering
    \includegraphics[width=0.95\linewidth]{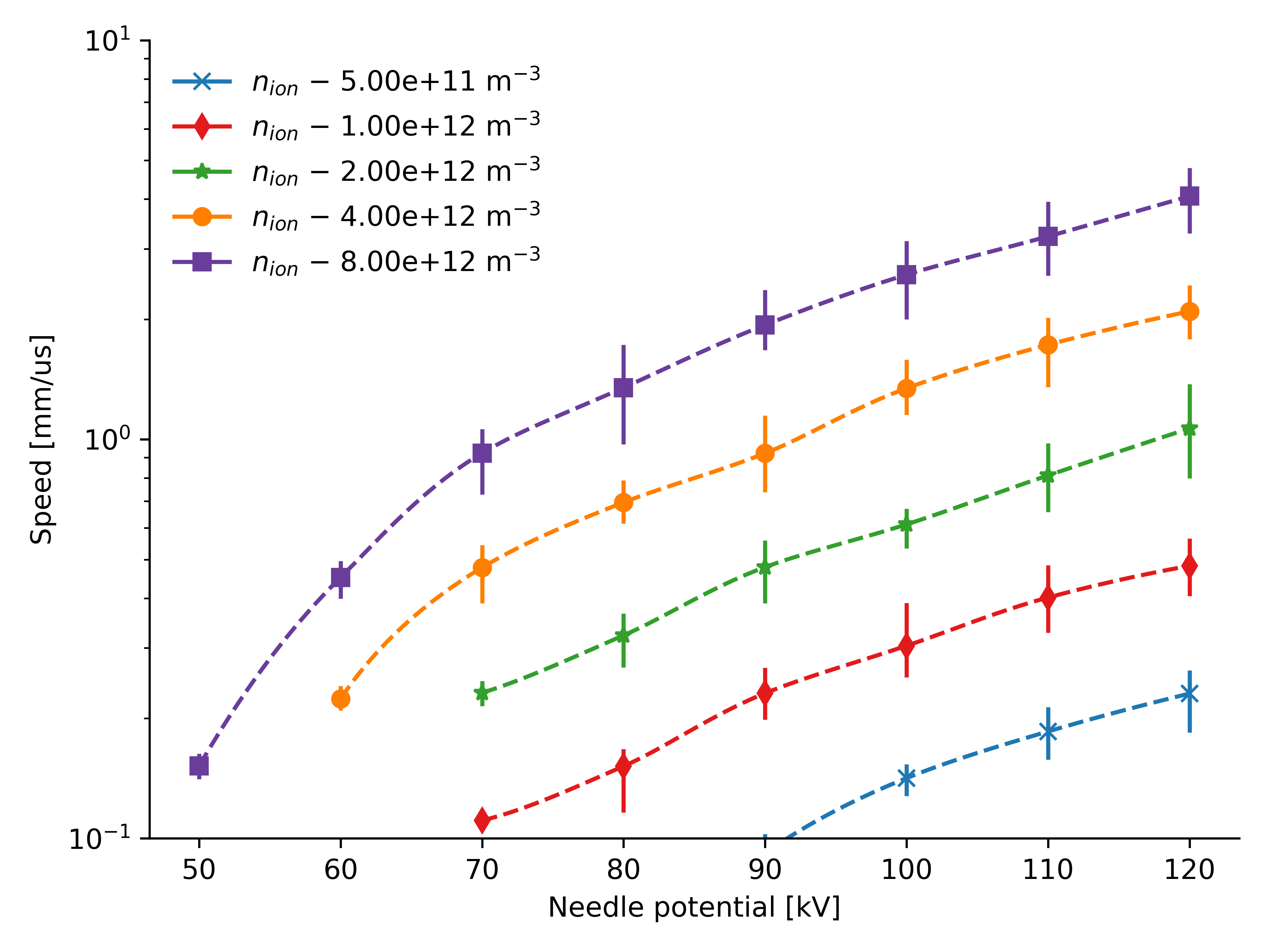}
    \caption{
        The effect of seed concentration $n_'ion'$
        on the average streamer propagation speed
        for the middle 50~\% of the gap.
        Streamers terminated in the first 25~\% of the gap are excluded.
        The default concentration is about \SI{2e12}{\metre^{-3}}. 
        The dashed lines are interpolated to the average,
        and the bars covers the minimum and maximum values.
        }
    \label{fig:zapsim_stat_lgy_v_psm_sc(c)}
\end{figure}

\begin{figure}[t!]
    \centering
    \includegraphics[width=0.95\linewidth]{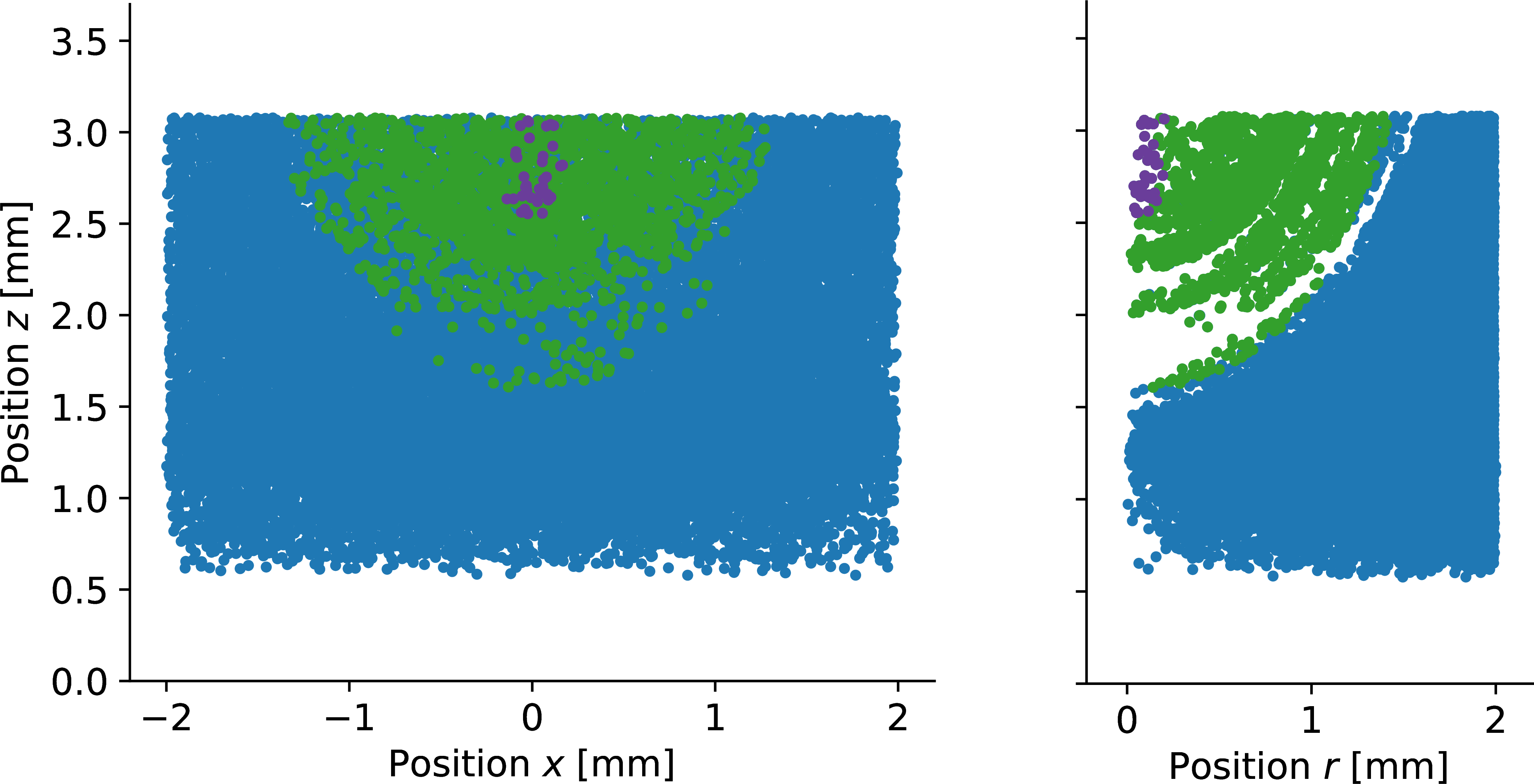}
    \caption{
        Streamer stop after sweep-out of too many electrons
        at $\SI{90}{\kV}$ and $E_'d' = \SI{15}{\mega\volt\per\meter}$.
        $xz$- and $rz$-projection where
        each seed is represented by a dot;
        anion (blue),
        electrons (green),
        and
        avalanches (purple).
    }
    \label{fig:seed_starvation}
\end{figure}

\begin{figure}[t!]
    \centering
    \includegraphics[width=0.95\linewidth]{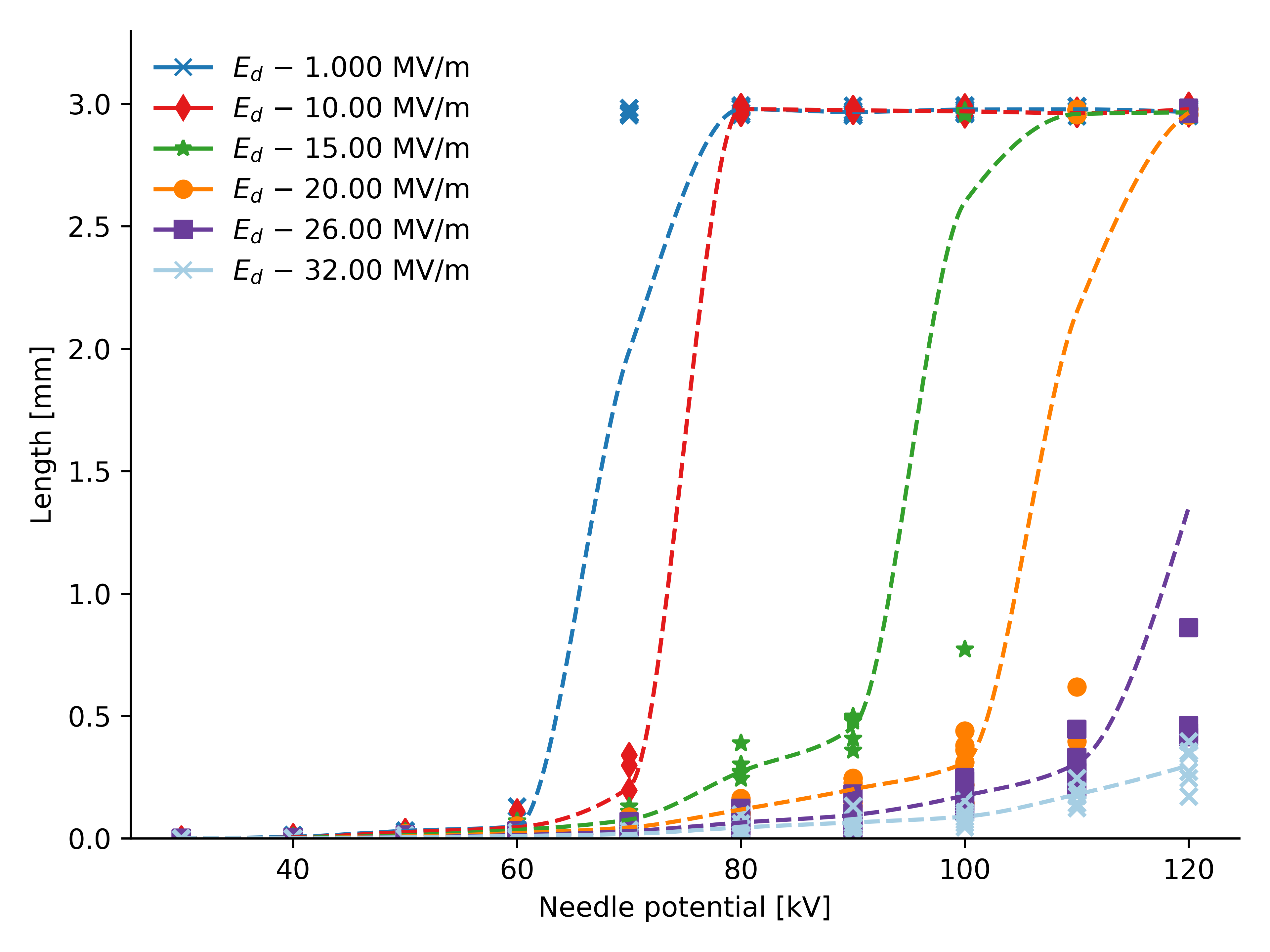}
    \caption{
        Streamer propagation length as a function of needle potential
        and electron detachment threshold $E_'d'$.
        Each marker is a simulation
        and the dotted lines are interpolated to the average.
        }
    \label{fig:zapsim_stat_v_ls_let(c)_mod}
\end{figure}

\begin{figure}[t!]
    \centering
    \includegraphics[width=0.95\linewidth]{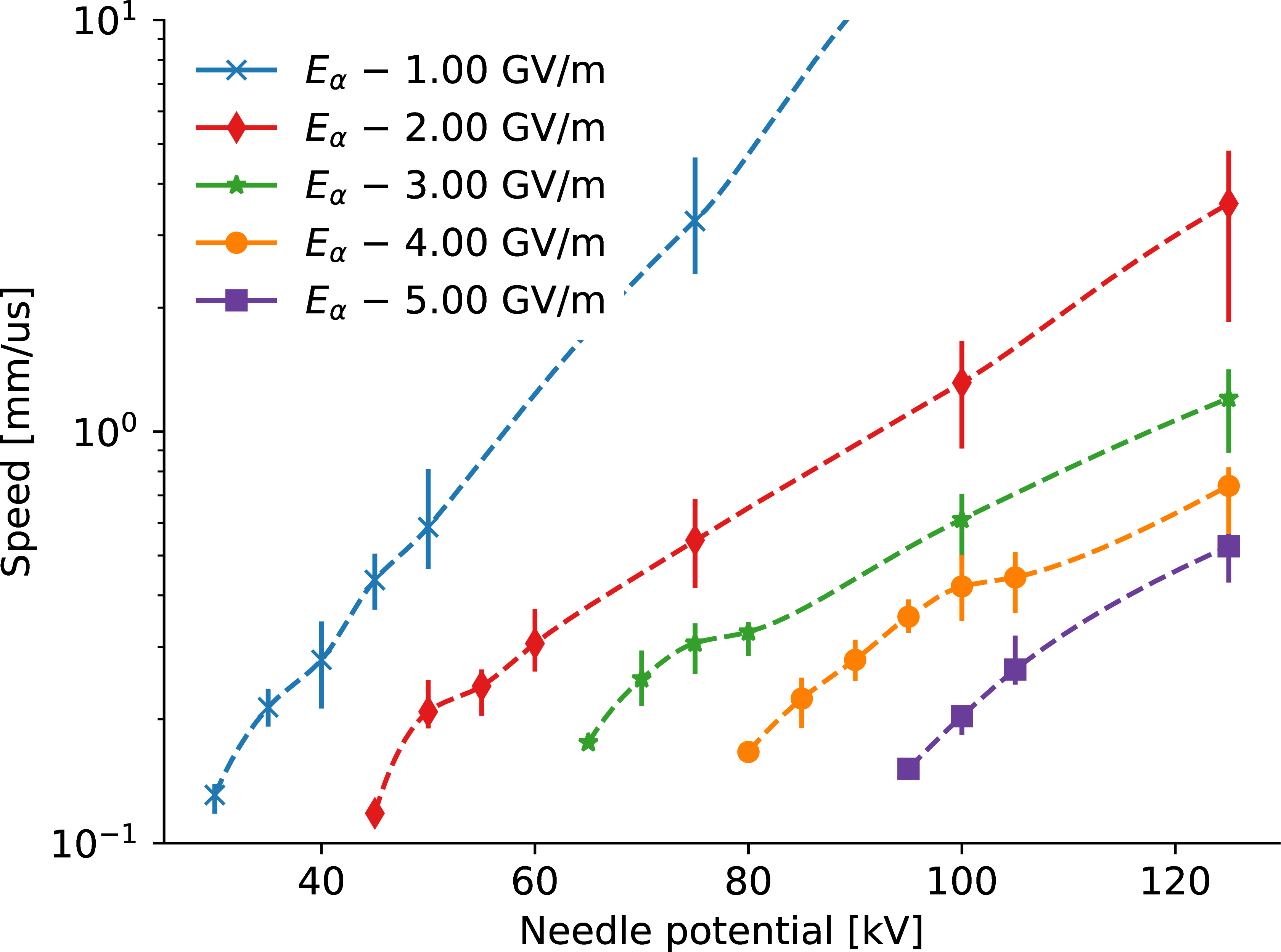}
    \caption{
        The effect of $E_\alpha$
        on the average streamer propagation speed
        for the middle 50~\% of the gap.
        The dashed lines are interpolated to the average,
        and the bars covers the minimum and maximum values.
        }
    \label{fig:zapsim_stat_lgy_v_psm_Ea(c)}
\end{figure}

\begin{figure}[t!]
    \centering
    \includegraphics[width=0.95\linewidth]{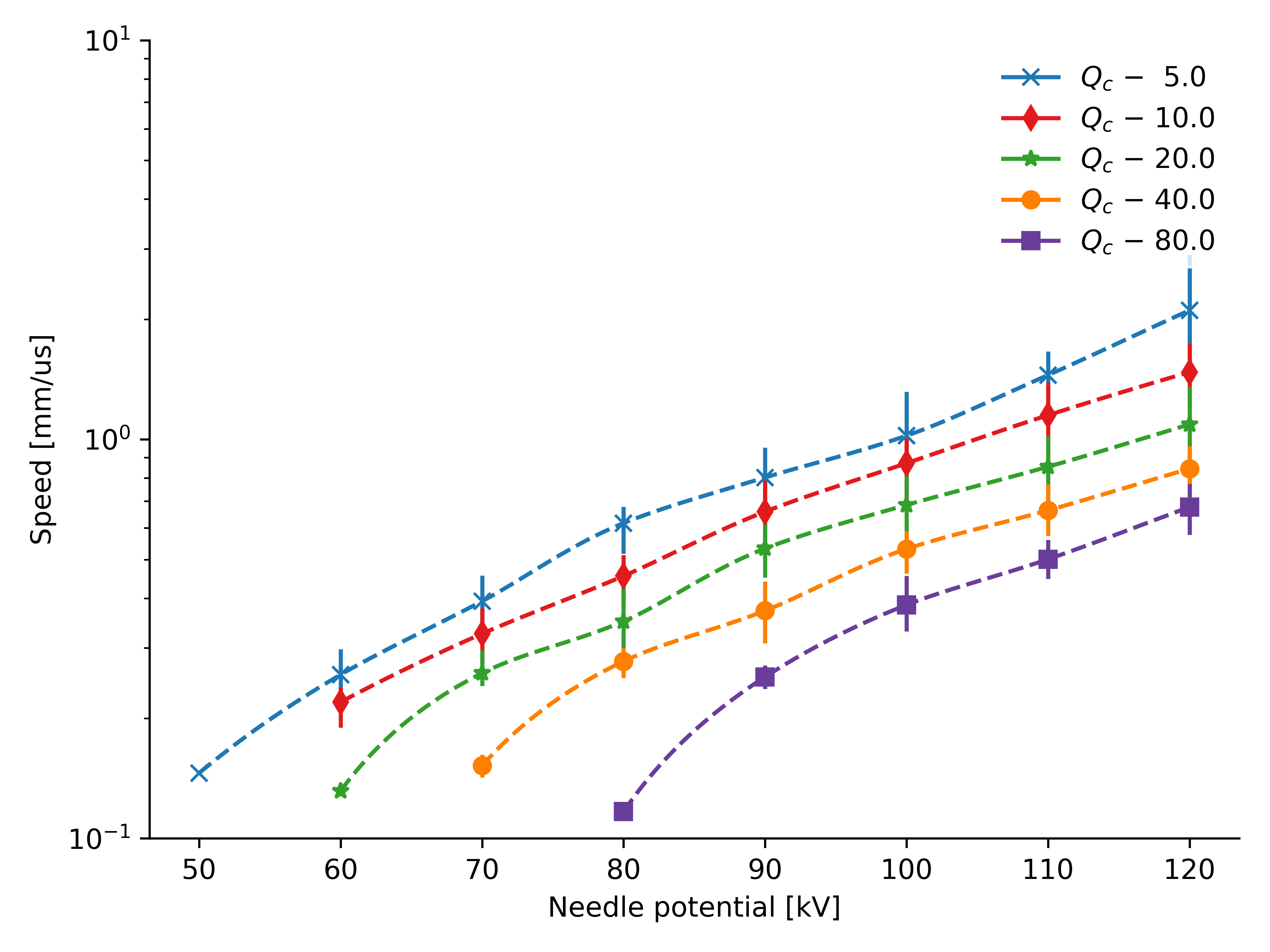}
    \caption{
        The effect of $Q_'c'$
        on the average streamer propagation speed
        for the middle 50~\% of the gap.
        The dashed lines are interpolated to the average,
        and the bars covers the minimum and maximum values.
        }
    \label{fig:zapsim_stat_lgy_v_psm_qc(c)}
\end{figure}

The avalanche mechanism is the most important part of the model.
For this reason,
parameters relevant to the avalanche growth,
given in \cref{eq:alpha,eq:Q_e},
are especially important.
To get an avalanche,
however,
a seed electron is needed.
A doubling of the concentration of seeds $n_'ion'$,
gives about a doubling in the propagation speed,
as seen in \cref{fig:zapsim_stat_lgy_v_psm_sc(c)}.
The figure shows the average speed for the mid 50~\% of the gap,
that is for a position from \SIrange{0.75}{2.25}{\mm}.
Since streamers terminated in the first quarter of the gap are not shown,
the figure also indicates that
the breakdown voltage is dependent on $n_'ion'$,
as increasing $n_'ion'$ allows propagation at lower voltages.
The streamer is represented by one or more heads,
and propagates as new heads are added in front of current heads.
As such, the leading head moves in a series of discrete ``jumps''.
The average streamer head jump length seems independent of $n_'ion'$,
indicating that the linear increase in propagation speed
is caused by a reduction in the time required
for an electron to become a critical avalanche.
At $n_'ion' = \SI{2e12}{\metre^{-3}}$,
the average distance between seeds is \SI{79}{\micro\meter},
while the average jump length is about \SI{6}{\micro\meter},
so $n_'ion'$ would have to be increased by some orders of magnitude
to affect the streamer jump distance.
Inhomogeneities on the order of \SI{e11}{\metre^{-3}}
was introduced by~\cite{Jadidian2013}
to explain branching,
but this effect is not found here.
An upper estimate on the ions available can be calculated
from \cref{eq:n_ep}
by using $G_'ion'$ instead of $G_'free'$
when calculating $R_'e'$ in \cref{eq:R_e}
and using a low estimate of $k_'r' = \SI{e-3}{mm^2 V^{-1} s^{-1}}$~\citep{Denat1979,Jadidian2013},
yielding $n_'ion' = \SI{1.8e13}{\metre^{-3}}$
and an average distance of \SI{38}{\micro\meter} between seeds.
As such,
the simulations in \cref{fig:zapsim_stat_lgy_v_psm_sc(c)}
cover the most interesting range.

The baseline results in \cref{sec:results_baseline}
do not show any stopping of streamer propagation mid-gap.
The streamers either stop
within the first \SI{100}{\micro\meter}
or cause a breakdown.
This occurs when the supply of electrons is constant
and $E_'s'$ is too low to create a high voltage drop along the streamer.
Increasing the electron detachment threshold $E_'d'$
reduces the number of electrons available,
which in turn reduces the density of electrons
as electrons are swept out,
see \cref{fig:seed_starvation}.
This results in a negative feedback loop where
a lower density of electrons decreases the speed
(\cref{fig:zapsim_stat_lgy_v_psm_sc(c)})
and the decreased speed results in a lower rate of
ions turning into electrons.
The propagation length is shown as a function of
the needle potential and $E_'d'$ in
\cref{fig:zapsim_stat_v_ls_let(c)_mod}.
By considering $E_'d' = \SI{15}{\mega\volt\per\meter}$,
three different regimes is identified.
Up to \SI{70}{\kV},
a few avalanches may occur,
but then the propagation stops.
Above \SI{90}{\kV},
the propagation is fast enough to provide
a stable rate of new electrons,
enabling the propagation to continue.
In between,
the initial electrons allow the streamer to propagate,
but the electron density is decreasing
and the streamer eventually stops.

The electric field is important for electron movement and multiplication,
and $E_\alpha$ in \cref{eq:alpha} is therefore an important parameter.
The strong influence of $E_\alpha$ is seen in
\cref{fig:zapsim_stat_lgy_v_psm_Ea(c)},
where the propagation speed may increase by an order of magnitude
when $E_\alpha$ is reduced by 50~\%.
This makes sense
as $E_\alpha$ enters exponentially
in \cref{eq:alpha}.
The propagation speed of 2nd mode streamers
is weakly dependent on the applied voltage~\citep{Lesaint2016},
however,
for $E_\alpha = \SI{1}{\giga\volt\per\meter}$
in \cref{fig:zapsim_stat_lgy_v_psm_Ea(c)},
the dependence is much stronger than for the other values.
Reducing $E_\alpha$ facilitates streamer propagation
and the breakdown voltage is thus strongly influenced.
Both $E_\alpha$ and $\alpha_'m'$
are based on experimental results,
and are very important to the model.
Instead of varying $\alpha_'m'$,
however,
the Meek-constant $Q_'c'$ is varied.
From \cref{eq:alpha,eq:DQ,eq:Q_e}, it is clear that
the avalanche size $Q_'e'$ is linearly dependent on $\alpha_'m'$,
which implies that doubling $Q_'c'$
has the same effect as halving $\alpha_'m'$.
The speed is not as affected by $Q_'c'$ as intuitively expected,
see \cref{fig:zapsim_stat_lgy_v_psm_qc(c)},
and changing $Q_'c'$ by a factor of 4
only changes the speed by a factor of 2.
However, $Q_'c'$ cannot change much
before the simulation becomes unphysical.
For instance,
consider a conducting sphere of $r = \SI{6}{\micro\meter}$
with a charge $q = \exp(Q_'c')$.
The electric field at the surface is
\begin{align}
    E = \frac{e q}{4 \pi \epsilon r^2} \,,
\end{align}
where $e$ is the electron charge and $\epsilon$ is the permittivity.
For $Q_'c'$ equal 15, 20, and 25,
the electric field becomes
\SI{6.5e7}{\volt \per \metre},
\SI{9.7e9}{\volt \per \metre}, and
\SI{1.4e12}{\volt \per \metre}, respectively.
Increasing $Q_'c'$ by a little gives too high fields,
and a decrease results in low fields.
This can,
however,
be ``fixed'' by changing the radius.
For instance, $Q_'c' = 15$ and $r = \SI{1}{\micro\meter}$,
results in \SI{2.4e9}{\volt \per \metre},
which is more reasonable.
To consider the electron avalanche as a charged sphere
is of course a simplification,
but the majority of the charge does build up
over a length of some \si{\micro\meter},
and this is also the size used for the streamer heads,
which makes the analogy reasonable.
While it would seem like increasing $Q_'c'$
does not make sense,
one should remember that
it actually has the same effect on the model
as decreasing $\alpha_'m'$,
and the value of that parameter is not certain.
For instance, according to \cite{Haidara1991},
$\alpha_'m' = \SI{200}{\per \micro\meter}$,
but \cite{Naidis2015tdei} finds
$\alpha_'m' = \SI{130}{\per \micro\meter}$,
however,
the latter study also finds
$E_\alpha = \SI{1.9}{\giga\volt \per \metre}$,
and changing this parameter has a big impact on the model,
as discussed above.

%


\subsection{Effect of streamer parameters}

\begin{figure}[t!]
    \centering
    \includegraphics[width=0.95\linewidth]{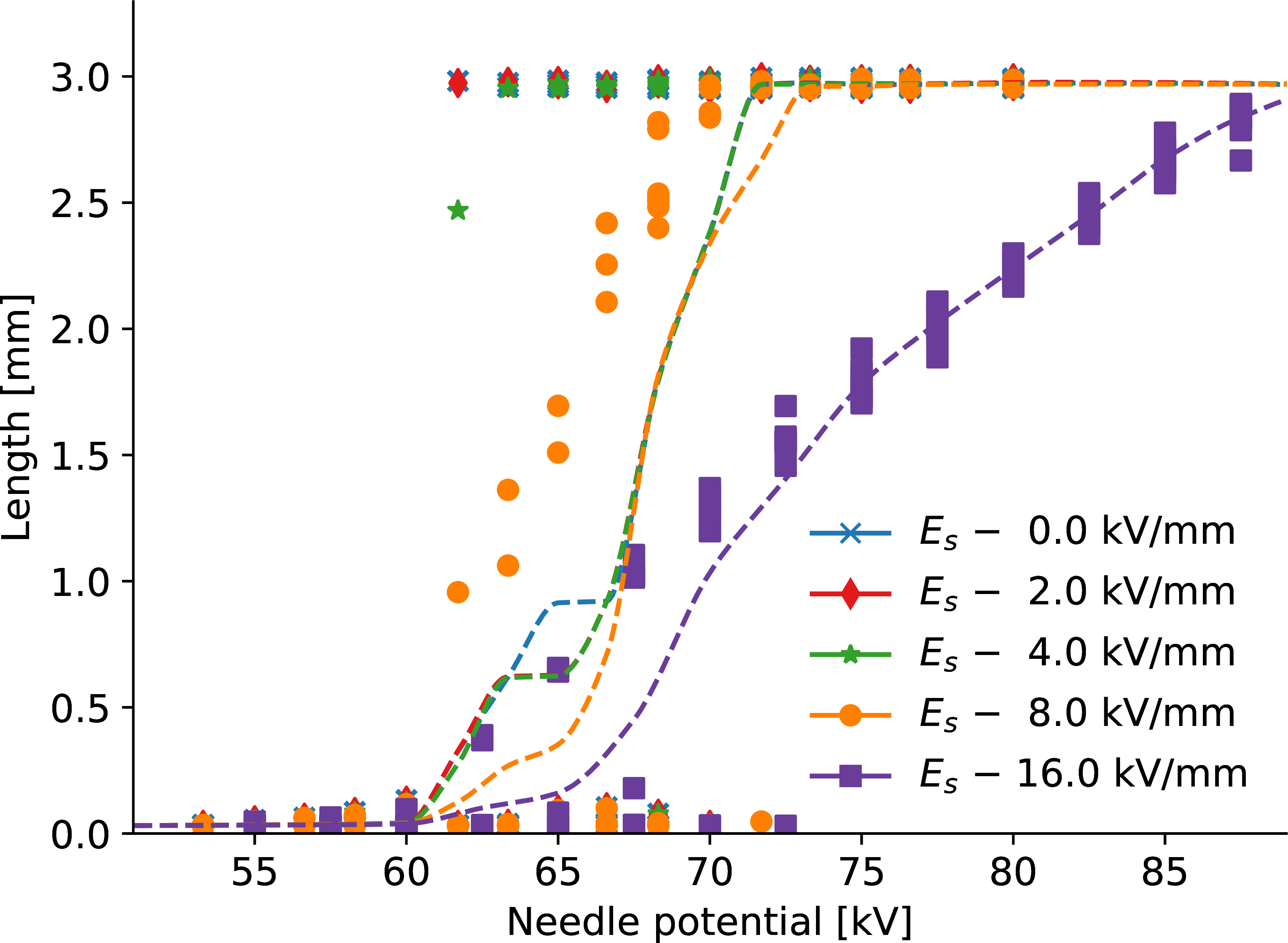}
    \caption{
        Streamer propagation length as a function of needle potential
        and electric field in streamer channel $E_'s'$.
        Each marker is a simulation
        and the dotted lines are interpolated to the average.
        Note that up to \SI{8}{\kV\per\metre},
        the results overlap to a high degree.
        }
    \label{fig:zapsim_stat_v_ls_sug(c)}
\end{figure}

\begin{figure}[t]
    \centering
    \includegraphics[width=0.95\linewidth]{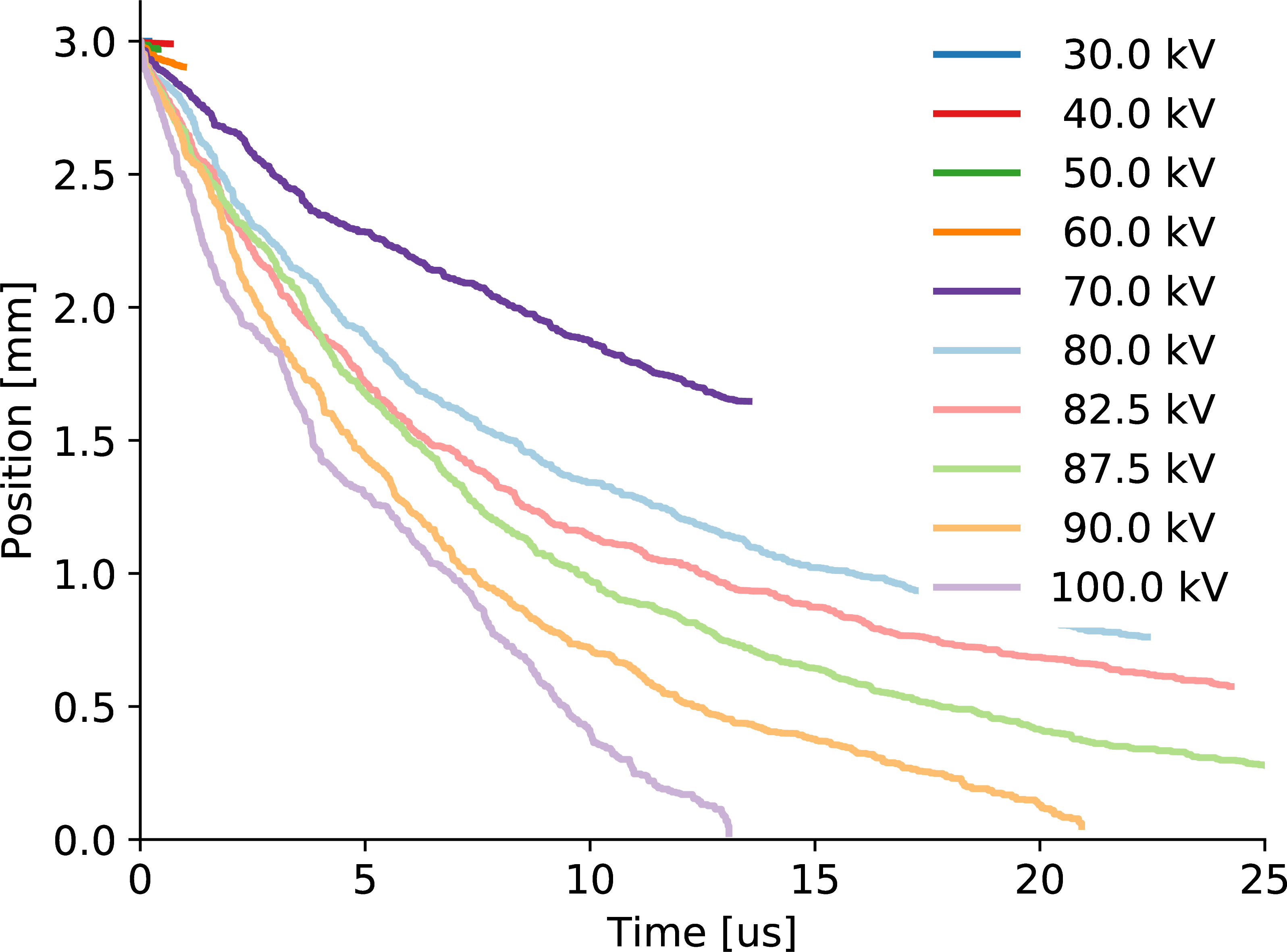}
    \caption{
        Streak plots of streamer leading head position,
        using $E_'s' = \SI{16}{\kV \per \mm}$,
        causing the streamers to slow down and sometimes stop.
        }
    \label{fig:multi_streak_zapsim_300_stat}
\end{figure}

\begin{figure}[t!]
    \centering
    \includegraphics[width=0.95\linewidth]{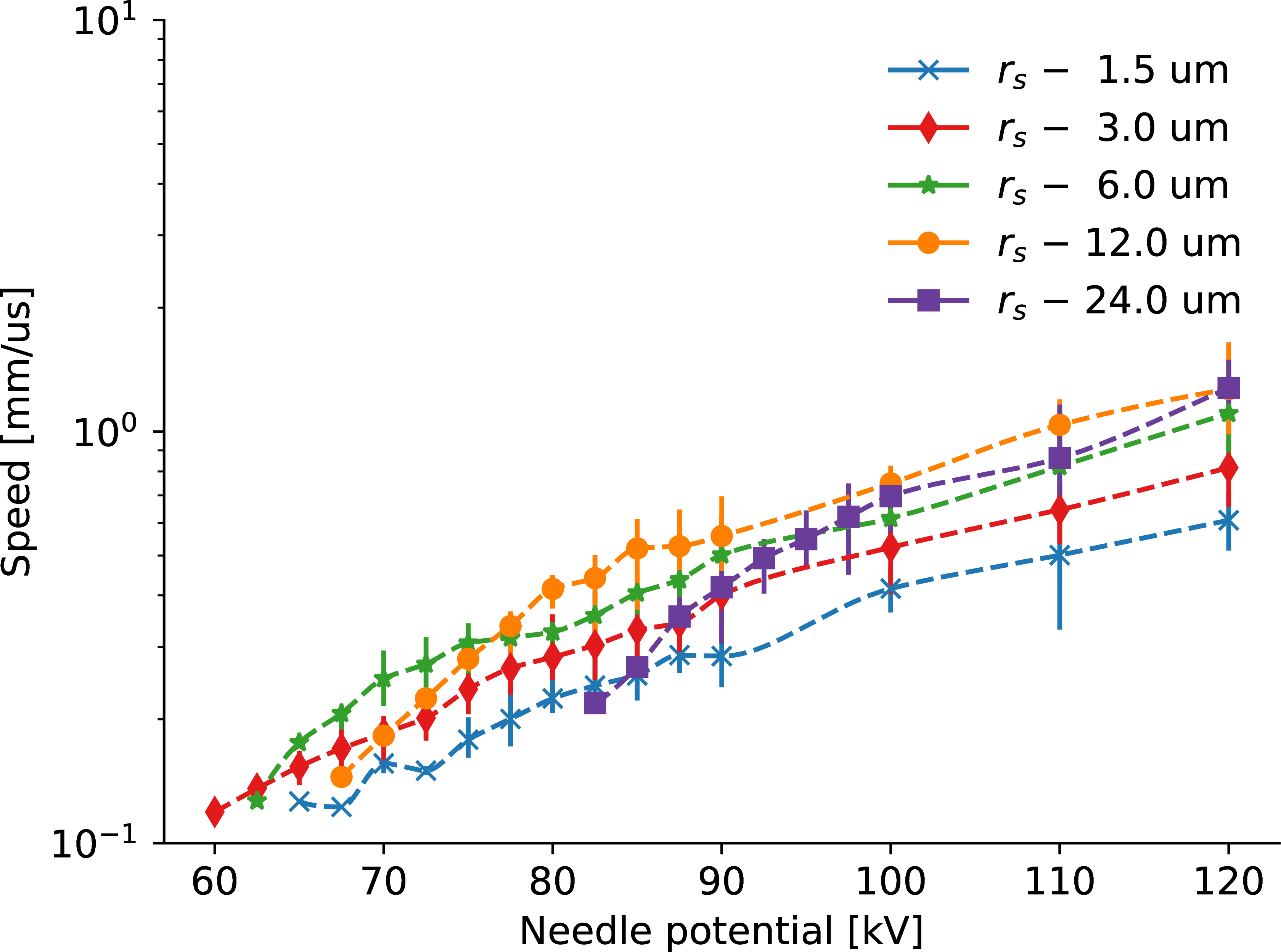}
    \caption{
        Streamer propagation speed for a series of different
        streamer head tip curvatures $r_'s'$.
        The dotted lines are interpolated to the average,
        and the bars covers the minimum and maximum values.
        }
    \label{fig:zapsim_stat_lgy_v_psm_srp(c)}
\end{figure}

\begin{figure}[t!]
    \centering
    \includegraphics[width=0.95\linewidth]{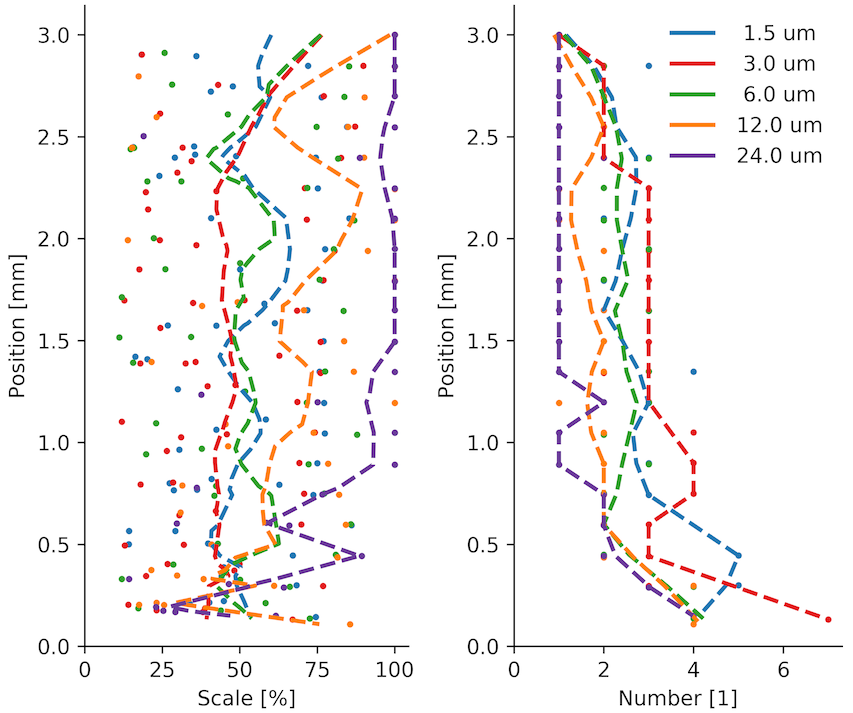}
    \caption{
        Actual streamer head scale $k_i$ (left)
        and total number of streamer heads (right)
        at \SI{100}{\kV}
        for a series of
        streamer head tip curvatures $r_'s'$.
        Data are taken every 5~\% of the gap.
        The dashed lines are moving averages
        calculated by loess-regression~\citep{Cleveland1979}.
        }
    \label{fig:multi_tipsno_tipscale_zapsim_070_gp5}
\end{figure}

\begin{figure}[t!]
    \centering
    \includegraphics[width=0.95\linewidth]{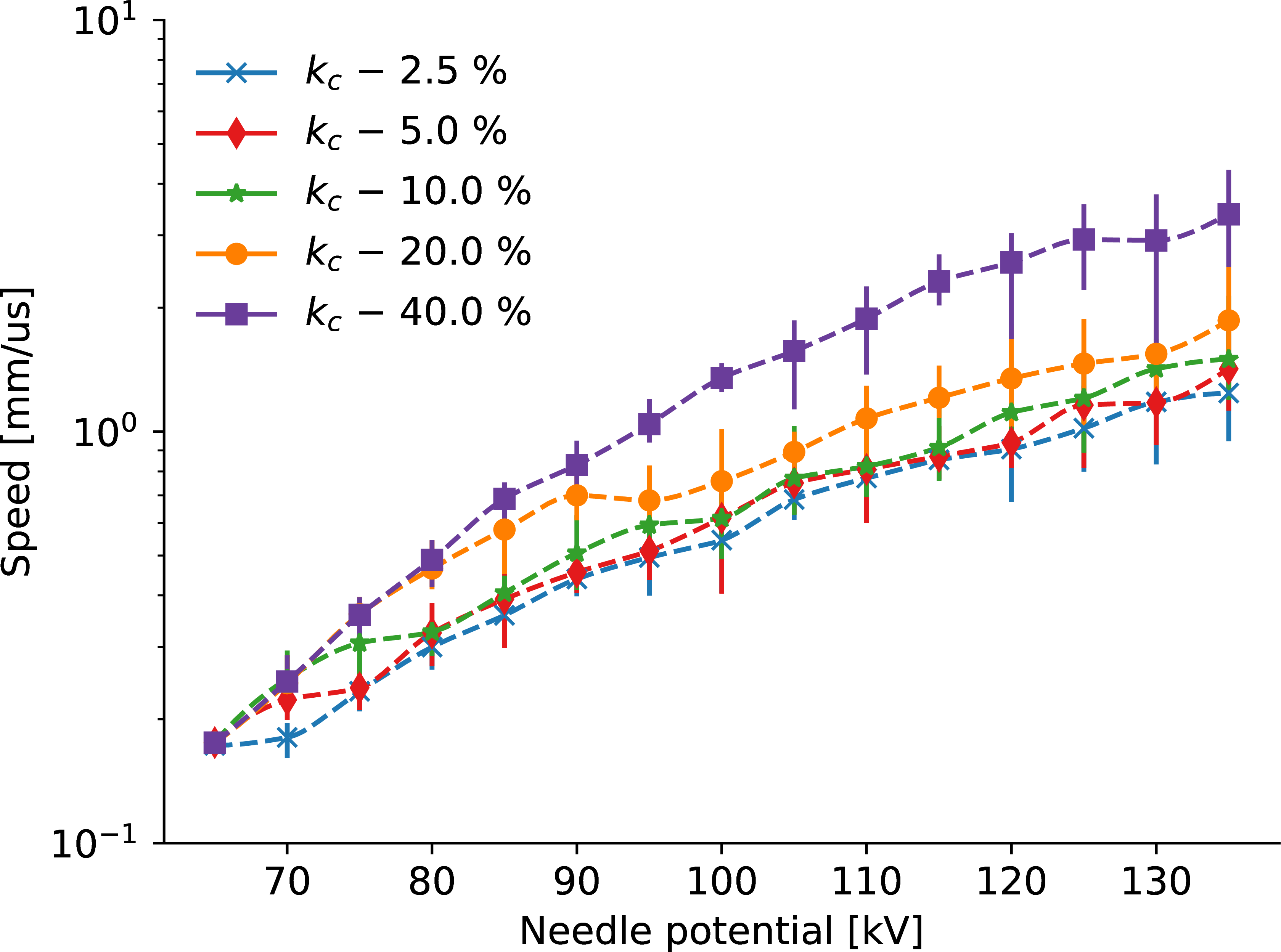}
    \caption{
        The effect of streamer head scale threshold
        $k_'c'$ on the streamer propagation speed,
        calculated for the mid 50~\% of the gap.
        The dotted lines are interpolated to the average,
        and the bars covers the minimum and maximum values.
        }
    \label{fig:zapsim_stat_lgy_v_psm_sst(c)}
\end{figure}

\begin{figure}[t]
    \centering
    \includegraphics[width=0.95\linewidth]{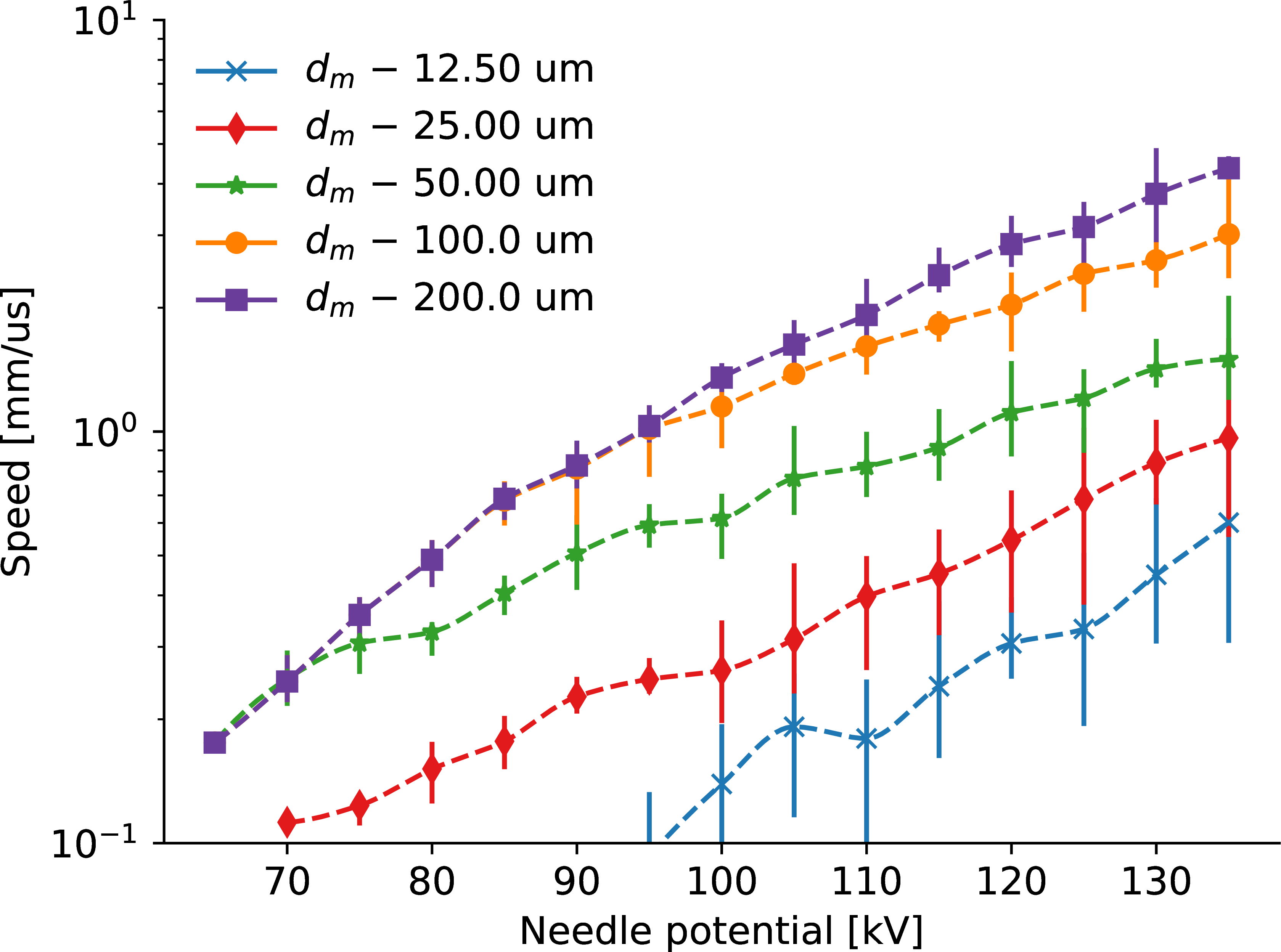}
    \caption{
        The effect of streamer merge distance
        $d_'m'$ on the streamer propagation speed,
        calculated for the mid 50~\% of the gap.
        The dotted lines are interpolated to the average,
        and the bars covers the minimum and maximum values.
        }
    \label{fig:zapsim_stat_lgy_v_psm_dm(c)}
\end{figure}

\begin{figure}[t]
    \centering
    \includegraphics[width=0.95\linewidth]{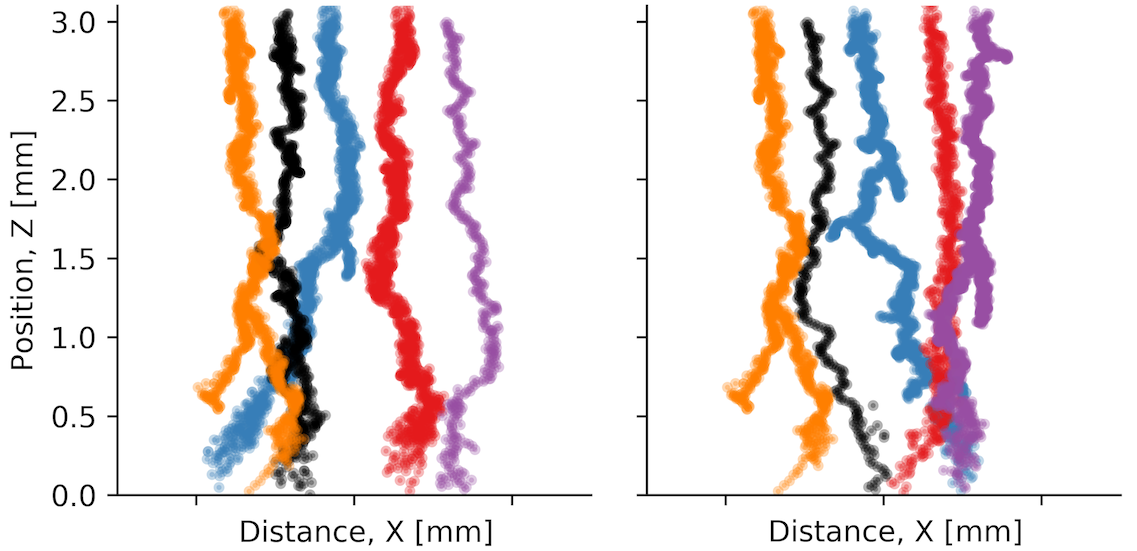}
    \caption{
        Streamer head positions for simulations at \SI{120}{\kV}.
        Variation of $k_'c'$ (left): 
        10~\% (yellow),
        20~\% (black),
        2.5~\% (blue),
        5.0~\% (red),
        and 40~\% (purple).
        Variation of $d_'m'$ (right): 
        \SI{50.0}{\micro\meter} (yellow),
        \SI{200}{\micro\meter} (black),
        \SI{25.0}{\micro\meter} (blue),
        \SI{100}{\micro\meter} (red),
        and \SI{12.5}{\micro\meter} (purple).
                }
    \label{fig:multi_trail_zapsim_090_stat_kc_sst_2x2}
\end{figure}

The streamer structure is responsible for
propagating the electric field
from the needle into the gap.
The electric field in the streamer channel $E_'s'$
gives a voltage drop from the needle to the streamer head.
The electric field in front of a streamer head
is also dependent on
the tip radius of curvature $r_'s'$
and the potential scaling of the streamer head $k_'i'$.
The scaling depends on the potential and position of all the streamer heads,
that is, the entire ``streamer''.
Both the streamer head merge distance $d_'m'$
and the potential shielding threshold $k_'c'$
may be important for the streamer configuration.

\Cref{fig:zapsim_stat_v_ls_let(c)_mod} demonstrates
streamers stopping as a result of
a reduction in the seed electron density,
however,
it is common to explain stopping as a result of
an electric field $E_'s'$ in the streamer channel
resulting in a lower field strength at the streamer head%
~\citep{Atten1993}.
A high $E_'s'$ is needed to affect the results
(see \cref{fig:zapsim_stat_v_ls_sug(c)}),
conversely, when $E_'s'$ is low,
the streamer either stops quickly
or causes a breakdown.
When $E_'s'$ is high,
the propagation speed is reduced
throughout the gap
and the propagation may stop somewhere in the gap,
see \cref{fig:multi_streak_zapsim_300_stat}
for $E_'s' = \SI{16}{\kV \per \mm}$,
which is in contrast to
\cref{fig:fig_kc_multi_streak_zapsim_930_stat}
for $E_'s' = \SI{2.0}{\kV \per \mm}$
where the streamers do not stop.
Both \cref{fig:zapsim_stat_v_ls_sug(c),fig:multi_streak_zapsim_300_stat}
indicate that $E_'s'$
is not important in the beginning of the propagation,
but becomes important when a streamer has reached some length.
When $E_'s' = \SI{8}{\kV \per \mm}$,
the potential is reduced by \SI{24}{\kV} across the gap,
but this effect is barely seen (\cref{fig:zapsim_stat_v_ls_sug(c)}),
since only a few streamers stop mid-gap.
However,
at $\SI{16}{\kV \per \mm}$
the effect is clearly present
as many of the streamers stop mid-gap.
Notice that at \SI{75}{\kV} to \SI{85}{\kV},
in \cref{fig:zapsim_stat_v_ls_sug(c)}
the average propagation length is increased
from about \SI{1.7}{\mm} to  \SI{2.6}{\mm},
giving an apparent electric field of only $\SI{11}{\kV \per \mm}$
and not \SI{16}{\kV \per \mm}.
This is perhaps an effect of the field increasing
as the gap is getting smaller.
Also,
actual experiments show stopping lengths
that are increasing linearly with voltage
in the first part of the gap,
followed by more scatter and superlinear behavior
towards the end of the gap%
~\citep{Lesaint1988,Lesaint1998,Ingebrigtsen2009}.
This behavior is not seen in \cref{fig:zapsim_stat_v_ls_sug(c)},
possibly because $E_'s'$ is kept constant in the simulations,
while it has been found to vary with applied voltage~\citep{Massala1998}.
Streamers are subject to re-illuminations,
associated with current pulses,
which could change the electric field in the streamer channel,
however,
the propagation of the streamer head
seems to be unaffected by these effects~\citep{Massala1998}.

The curvature radius $r_'s'$ of a streamer head
is an interesting parameter
since a sharper tip gives a higher field
and a larger volume where electron multiplication may occur.
Changing $r_'s'$ from \SI{1.5}{\micro\meter} to \SI{12}{\micro\meter}
only changes the speed by a factor of 2,
see \cref{fig:zapsim_stat_lgy_v_psm_srp(c)}.
Further increase to \SI{24}{\micro\meter}
decreases the speed,
and increases the breakdown voltage.
Simulations with smaller $r_'s'$ tend to have
more streamer heads, scaled to a lower potential,
than the simulations with a larger $r_'s'$,
indicated in \cref{fig:multi_tipsno_tipscale_zapsim_070_gp5},
although the effect is not visible for the smallest $r_'s'$
in that figure.
The increased number of streamer heads seems to act as
a regulating mechanism,
however, the number of branches is not increased,
but there are more streamer heads present simultaneously in the same branch.
This is similar to the situation in
\cref{fig:fig_kc_multi_trail_zapsim_930_stat,fig:fig_kc_multi_tipscale_tipsno_zapsim_930_gp1},
where an increased voltage does not increase the number of branches,
but instead increases the streamer thickness.

An increase in voltage
increases the speed
(\cref{fig:fig_kc_multi_speed_avg_zapsim_930_stat})
as well as the number of streamer heads,
while decreasing the scaling of the heads
as demonstrated in \cref{fig:fig_kc_multi_tipscale_tipsno_zapsim_930_gp1}.
The parameters $k_'c'$ and $d_'m'$ are used to remove streamer heads,
and therefore they could have a big impact on the model,
since the scaling,
which the electric field depends on,
is strongly dependent on the number of streamer heads
as well as their configuration.
Also, these parameters are purely a consequence of the model,
and do not have an origin in a physical property.
Simulation results for varying $k_'c'$ are found in
\cref{fig:zapsim_stat_lgy_v_psm_sst(c)}
and show that the propagation speed is
not that affected,
except for $k_'c' = 40~\%$.
This figure also indicates that
the breakdown voltage is unaffected,
since all the values of $k_'c'$ are present for all the voltages.
Setting $k_'c' = 40~\%$ restricts the streamer
to one head in most situations,
and keeping two heads in rare occasions,
which
gives an upper bound to the propagation speed
for each voltage.
From a computational point of view,
it is preferable to set $k_'c'$ high
as fewer streamer heads implies less calculation.
From a physical point of view,
however,
it does not make sense to just remove charges from the system,
so $k_'c'$ should be reasonably low.
According to \cref{fig:zapsim_stat_lgy_v_psm_sst(c)},
$k_'c'$ can be as high as 10~\%
without any particular impact on the results.

The influence of the streamer head merge distance $d_'m'$
is shown in \cref{fig:zapsim_stat_lgy_v_psm_dm(c)}.
For the lower values,
many streamer heads are present at the same time,
which in turn lowers the potential scaling of each head,
increases the breakdown voltage,
and moderates the propagation speed.
Increasing $d_'m'$ increases propagation speeds,
up to the limit where there is mainly just a single active streamer head.
\Cref{fig:zapsim_stat_lgy_v_psm_dm(c)}
also indicates that at low voltages,
the streamers propagate with a single head,
but when the voltage is increased
and more heads are possible,
the propagation speed is moderated.
As $d_'m'$ is increased,
the voltage needed to have several heads is also increased,
and the propagation speed is thus higher.
The set of streamers presented in
\cref{fig:multi_trail_zapsim_090_stat_kc_sst_2x2}
shows that the thickness of the streamers
is dependent on $k_'c'$ and $d_'m'$,
which is an indication of the number of streamer heads
present during propagation.
However, the figure does not indicate
a change in the number of major branches.


\subsection{Effect of additives}

\begin{figure}[t!]
    \centering
    \includegraphics[width=0.95\linewidth]{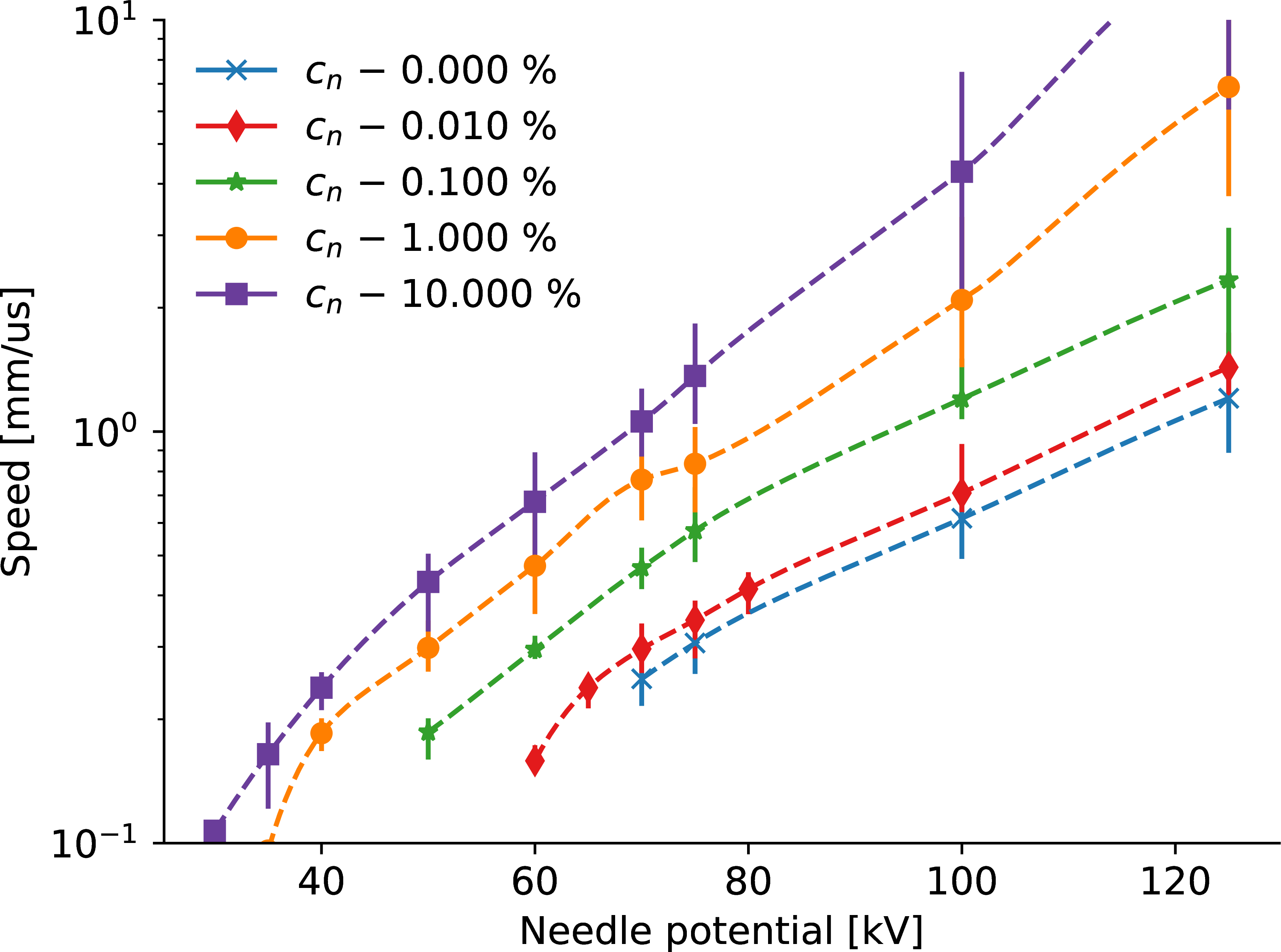}
    \caption{
        Streamer propagation speed for various fractions of added additive $c_'n'$.
        Average speed for the middle 50~\% of the gap.
        Each marker is a simulation
        and the dotted lines are interpolated to the average.
        }
    \label{fig:zapsim_stat_lgy_v_psm_cn(c)}
\end{figure}

Adding small amounts of an additive
increases the electron multiplication according to \cref{eq:alpha'}.
The effect should be similar to an increase of $\alpha_'m'$,
or a decrease in $Q_'c'$,
as discussed above
and shown in \cref{fig:zapsim_stat_lgy_v_psm_qc(c)}.
This is indeed the case,
the propagation speed increases
and the breakdown voltage
decreases with increasing content of
an additive with low ionization potential,
see \cref{fig:zapsim_stat_lgy_v_psm_cn(c)}.
When the liquid consists of
$c_'a,n' = 10~\%$ additive (mole fraction)
it cannot be argued to be a ``small amount'' of additive.
Even as little as 1~\% could be too much.
As mentioned in \cref{sec:additives},
an addition of just 0.1~\% increases the avalanche growth
by a factor of 6.9,
when using \cref{eq:alpha'} and the parameters in
\cref{tab:tab_param_phys}.
A decrease in breakdown voltage
and an increase in propagation speed
is also found in experiments
with low-IP additives~\citep{Devins1981,Lesaint2000,Ingebrigtsen2009},
however,
increased branching is also seen in the experiments
in contrast to the simulation results here.


\subsection{Increased speed and branching}\label{sec:naidis}

\begin{figure}[t!]
    \centering
    \includegraphics[width=0.95\linewidth]{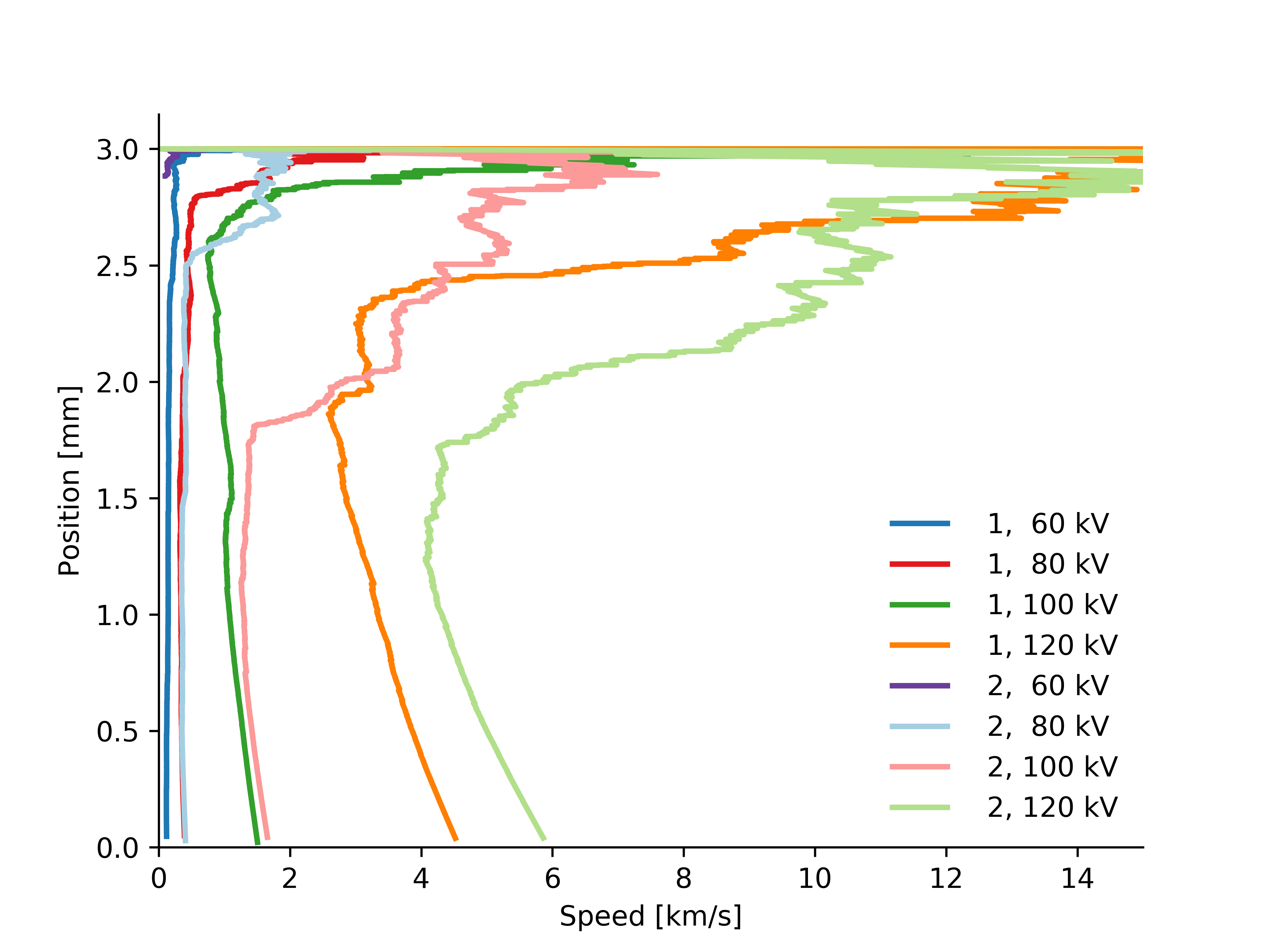}
    \caption{
        Streamer average speed versus leading head position.
        The simulations at the same voltage differ only by the
        initialization of the random number generator.
        }
    \label{fig:multi_speed_avg_zapsim_030_stat}
\end{figure}

\begin{figure}[t!]
    \centering
    \includegraphics[width=0.95\linewidth]{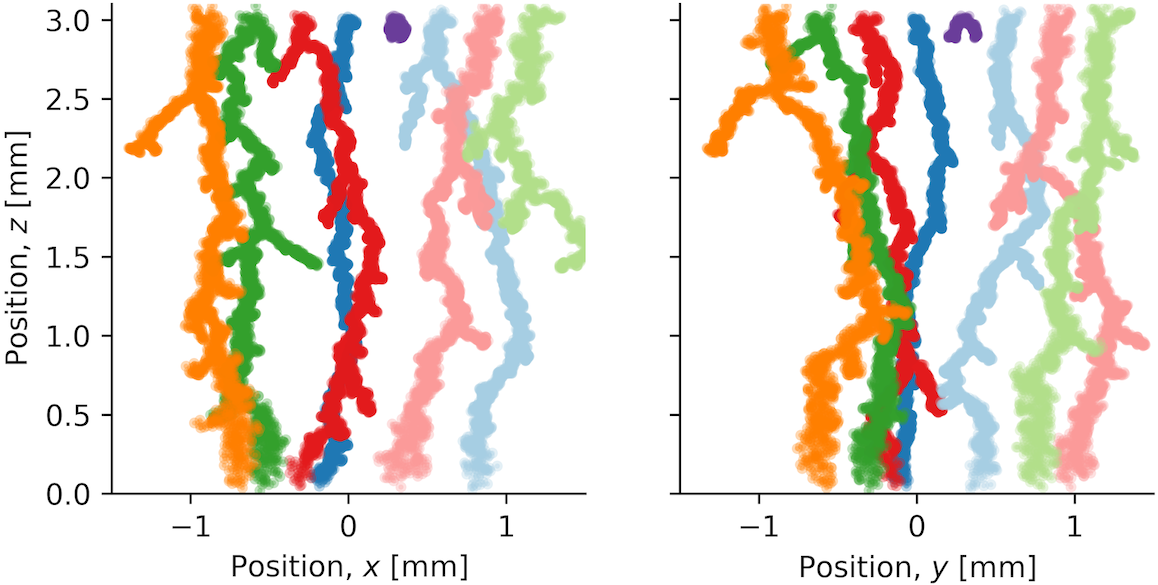}
    \caption{
        Streamer trails for a range of voltages,
        using the same colors as in \cref{fig:multi_speed_avg_zapsim_030_stat}.
        Each dot represents the position of a streamer head
        at some point of the propagation.
        }
    \label{fig:multi_trail_zapsim_030_stat_x4}
\end{figure}

\begin{figure}[t!]
    \centering
    \includegraphics[width=0.95\linewidth]{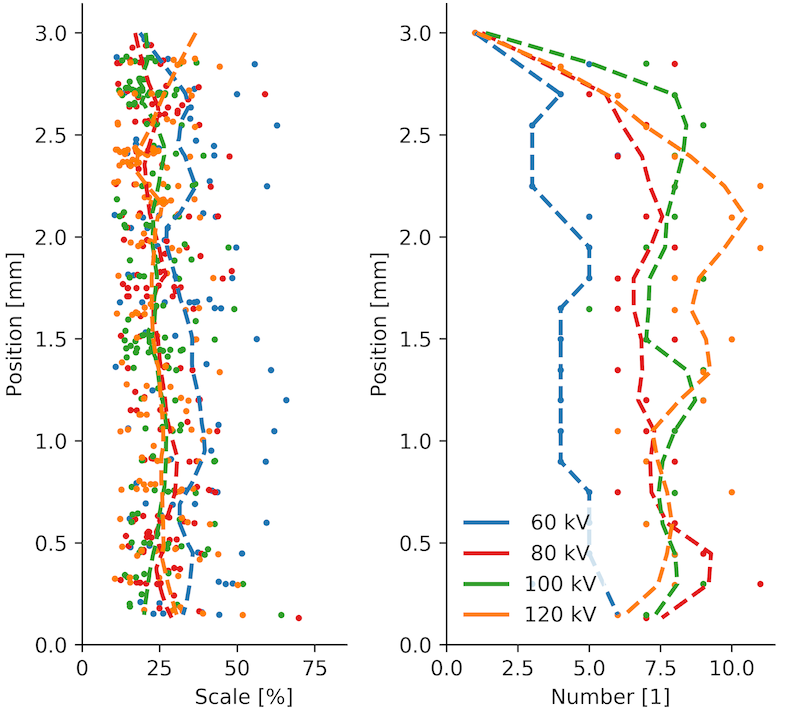}
    \caption{
        Actual streamer head scale $k_i$ (left)
        and total number of streamer heads (right).
        Data are taken every 5~\% of the gap.
        The dashed lines are moving averages.
        }
    \label{fig:multi_tipscale_tipsno_zapsim_030_gp5}
\end{figure}

\begin{figure}[t]
    \centering
    \includegraphics[width=0.95\linewidth]{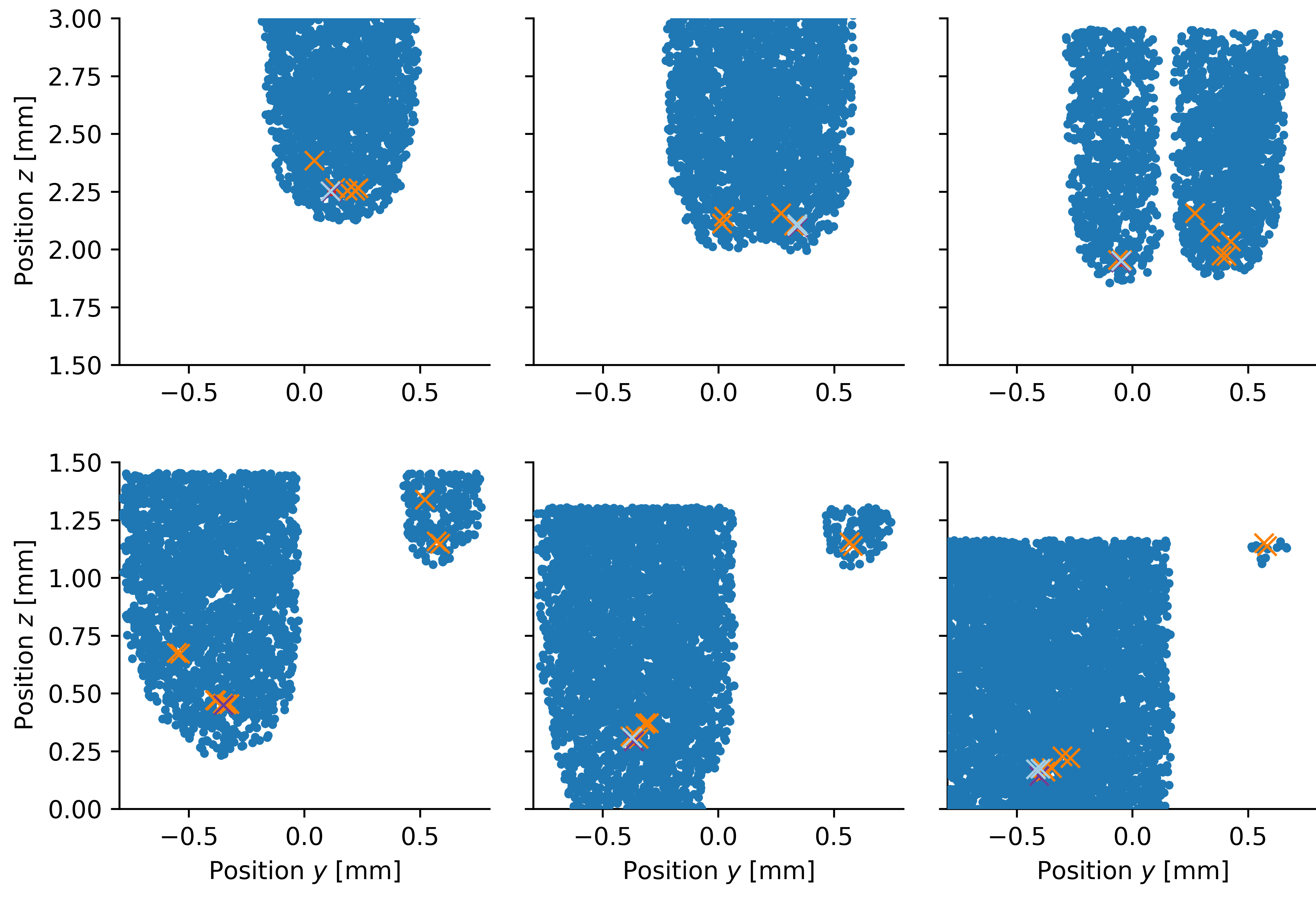}
    \caption{
        Detailed illustration of streamer branching.
        The electron avalanches are shown as blue dots,
        and the streamer head as crosses.
        The top three plots show streamer bifurcation
        in the start of the gap,
        while the bottom three plots
        show one propagating branch and another stopped branch
        at the end of the gap.
        }
    \label{fig:zapsim_053_gp5_branch}
\end{figure}

The above sections illustrate how the model behaves
and how it is affected by the various parameters.
In order to reduce the initiation voltage and
increase the propagation speed,
the avalanche parameters
are changed to
$E_\alpha = \SI{1.9}{\giga\volt\per\metre}$
and
$\alpha_'m' = \SI{130}{\per\micro\metre}$,
and the number of seeds is increased to
$c_'s' = \SI{8e12}{\metre^{-3}}$.
In addition, the merge distance is changed to
$d_'m' = \SI{12.5}{\micro\metre}$
and the streamer head tip radius to
$r_'s' = \SI{3}{\micro\metre}$
in order to facilitate branching.
Also, using
$E_'s' = \SI{8}{\kV\per\milli\metre}$
should be enough for some of the streamers to stop mid-gap.
Most of the predicted results are found:
the speed in
\cref{fig:multi_speed_avg_zapsim_030_stat}
is clearly increased compared to
\cref{fig:fig_kc_multi_speed_avg_zapsim_930_stat},
the amount of small branches is larger in
\cref{fig:multi_trail_zapsim_030_stat_x4}
than in
\cref{fig:fig_kc_multi_trail_zapsim_930_stat},
and
the decrease in streamer head scaling
and increase in streamer head number
is seen by comparing
\cref{fig:multi_tipscale_tipsno_zapsim_030_gp5}
to
\cref{fig:fig_kc_multi_tipscale_tipsno_zapsim_930_gp1}.
The propagation voltage is somewhat
lower than the base case,
around \SI{60}{\kV}.
The streamer propagation begins at high speed,
then slows down towards the middle of the gap,
before the speed increases towards the end of the gap,
see \cref{fig:multi_speed_avg_zapsim_030_stat}.
This change does not seem to be
correlated to the number of streamer heads,
which is fairly constant for most of the propagation
(\cref{fig:multi_tipscale_tipsno_zapsim_030_gp5}).
Branching may have an effect,
and streamer branching is illustrated in
\cref{fig:zapsim_053_gp5_branch},
showing 6 snapshots of a single simulation.
As the streamer splits into two major branches,
the number of electron avalanches
surrounding the streamer heads
decreases.
The branches propagate at different speeds,
and the faster one gains a higher potential
and thus creates more electron avalanches.
As the two branches
approaches the end of the gap,
one gains speed,
while the other one stops.

}

%


\section{Discussion of the model}\label{sec:discussion}{

\begin{figure}[t]
    \centering
    \includegraphics[width=0.95\linewidth]{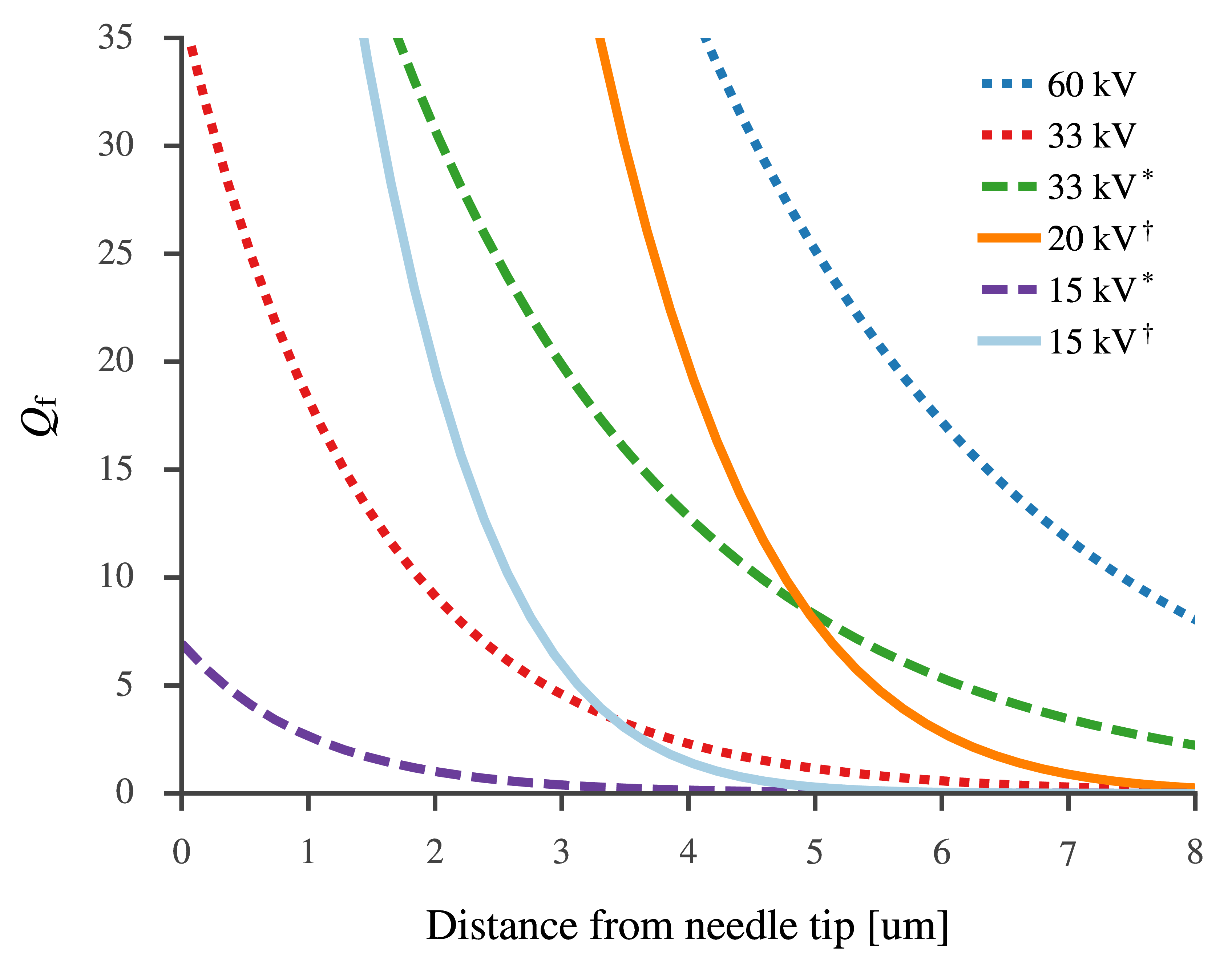}
    \caption{
        Maximum avalanche size at a distance from the needle tip.
        The unmarked lines use the same values
        as the baseline simulations from~\cite{Haidara1991},
        the $*$ indicates parameter values from~\cite{Naidis2015tdei}
        as used in \cref{sec:naidis},
        and
        the $\dagger$ indicates formulation and
        parameter values from~\cite{Atrazhev1991}.
        }
    \label{fig:alpha4_test_t2}
\end{figure}

\begin{figure}[t!]
    \centering
    \includegraphics[width=0.95\linewidth]{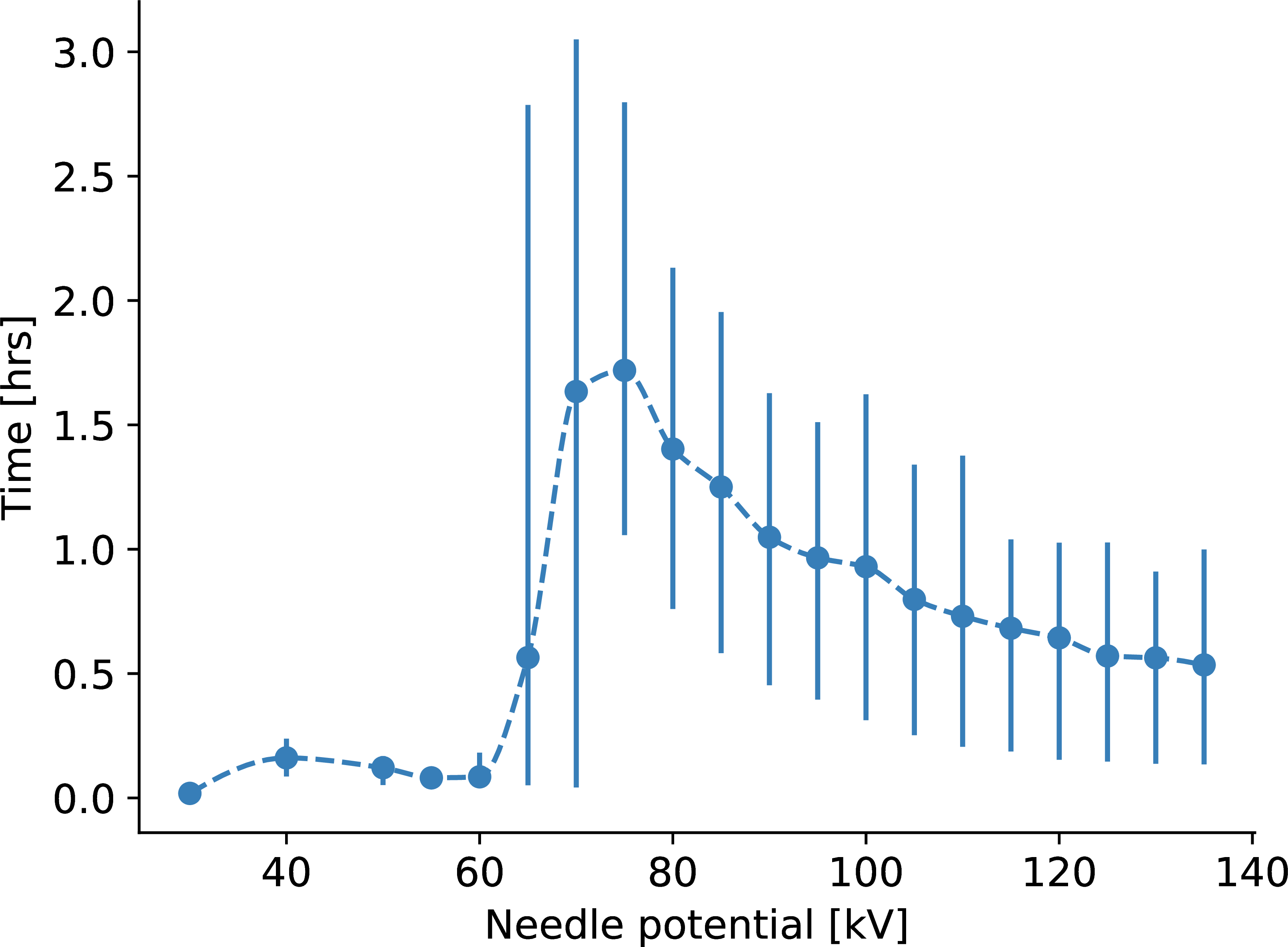}
    \caption{
        Total computational time for the simulations
        shown in \cref{fig:zapsim_stat_lgy_v_psm_sst(c)}
        Streamers that terminate in the beginning of the gap
        require little time,
        while streamers that slowly bridges the gap
        requires the most computational time.
        }
    \label{fig:zapsim_stat_v_ct_sst(c)_mod}
\end{figure}

Using a Laplace field is of course a simplification
compared to a Poisson field.
In fact, neither positive nor negative charges
are accounted for in the model.
The potential is simply calculated by assuming
a constant field in the streamer channel,
and then superimposing the streamer heads.
Including the charge of the avalanches
and the ions left behind
could improve the model.
For the needle and the streamer heads,
using a space charge limited field (SCLF)~\citep{Hibma1986,Boggs2005}
would provide a more physically correct field distribution,
but would also
increase the computational requirements drastically.
Using an SCLF rather than a Laplace field,
gives a reduction of the electric field
where the field is the strongest,
since the maximum field is limited~\citep{Hibma1986},
with a corresponding increase everywhere else.
The SCLF is time-dependent~\citep{Boggs2005},
and the effect increases with time
until an steady-state is obtained.
The overall effect on the model
would be an increase in average jump length,
as most jumps would be longer
and the shortest ones would not occur.
While an SCLF can give more accurate results for slow streamers,
a Laplace field could be good enough for fast streamers,
since the SCLF-region expands at some finite speed.
However,
the avalanche parameters in \cref{eq:alpha}
were estimated using a Laplace field,
so the current model is internally consistent.

The inception of 2nd mode streamers has been estimated to
somewhat less than \SI{15}{\kV} for cyclohexane~\citep{Gournay1993},
however, for a propagating 2nd mode streamer,
\SI{33}{\kV} was found
for a \SI{10}{\milli\meter} gap~\citep{Ingebrigtsen2009}.
Since the model uses this as a criterion for inception
(getting a critical avalanche, but no movement),
a high propagation voltage
is actually to be expected.
This is well illustrated
by the maximum avalanche size
in \cref{fig:alpha4_test_t2},
obtained by integration of $\alpha$.
Streamer propagation is possible when $Q_'f' > Q_'c'$
(cf. \cref{eq:Q_e}).
The baseline simulations
are performed inserting parameters from~\cite{Haidara1991}
in \cref{eq:alpha} to calculate $\alpha$.
At \SI{33}{\kV},
the maximum possible
streamer jump is less than a \si{\micro\meter},
however,
at \SI{60}{\kV} (the breakdown voltage),
the value is increased to about \SI{6}{\micro\meter},
possibly indicating that a strong field
is needed over some distance.
Changing to parameters from~\cite{Naidis2015tdei}
decreases the propagation voltage by
increasing the possible jump length,
however,
the decrease is not enough to enable
for inception of 2nd mode streamers at \SI{15}{\kV}.
As such, \cref{fig:alpha4_test_t2}
indicates that streamer inception at \SI{15}{\kV}
is not possible with this model
when considering a Laplace field,
calculating the electron multiplication with \cref{eq:alpha},
and using the Townsend--Meek criterion for inception of 2nd mode streamers.
Using the parameters of either~\cite{Haidara1991} or~\cite{Naidis2015tdei}
gives too low avalanche size.
According to~\cite{Atrazhev1991},
the correct way of calculating electron multiplication
in a dense medium is
\begin{equation}
    \alpha = \frac{3 I E_\nu^2}{e E} \exp \left({- \frac{E_\nu^2}{E^2}} \right)
    \label{eq_alpha_atrazhev}
\end{equation}
where
$I$ is the ionization potential,
$e$ is the electron charge,
and
$E_\nu$ is given by properties of the liquid.
With this formulation,
electron multiplication
is more dependent on the electric field,
implying that the electron avalanches
become shorter, are closer to the streamer heads,
and grow faster where the field is strong,
which is illustrated in \cref{fig:alpha4_test_t2}
using values for n-hexane~\citep{Atrazhev1991}.

The propagation velocity is somewhat low,
which is to be expected since the inception voltage is too high.
Changing parameters to values that lowers the inception voltage
also increases the speed at a given voltage.
As mentioned,
the speed is proportional to the electron mobility,
and it is the low-field mobility that has been used.
For low-mobility liquids, such as cyclohexane,
the mobility is expected to have a superlinear dependence on the electric field%
~\citep{Schmidt1977,Schmidt1984}.
For this reason,
one study multiplies the mobility by 2.5,
to make it similar to the gas phase mobility~\citep{Naidis2015jpd},
which would increase the streamer propagation speed by the same factor.
Conversely,
limitations to the maximum
speed of electrons have been introduced~\citep{Jadidian2014},
which would effectively control the maximum speed
of a streamer branch.
The speed is also proportional to the concentration of seeds
(see \cref{fig:zapsim_stat_lgy_v_psm_sc(c)}),
which was calculated from the low-field conductivity
of the liquid (see \cref{eq:n_ion}).
However,
for breakdown in non-polar liquids,
the conductivity is not important~\citep{Lesaint2016},
and hence, it seems unreasonable
for this parameter to be as important as demonstrated here.
The equilibrium density of ions can also be calculated
based on cosmic radiation~\cref{eq:N_e},
but obtaining $>\SI{e11}{m^{-3}}$ ions,
when the production is $\sim \SI{e8}{m^{-3} s^{-1}}$,
implies that a long time is needed.
It is therefore an approximation to simulate
a situation where this density is kept constant.
By changing the simulation conditions such that
all the gap is included in the ROI
and such that seeds are not replaced,
it can be verified that
the seeds present at the beginning of the experiment is not enough.
They are swept out very fast if they are electrons
and not ions.
Increasing $E_'d'$
so that most seeds remain as anions
changes this
by allowing the low-mobility anions to live longer
before entering the high-field area
and ionize into molecules and electrons.
Even so,
it seems clear that some mechanism
for generation of new seeds is warranted.
New seeds could be generated in the high-field areas,
and near the electrodes.
The Zener model~\citep{Zener1934} (field-ionization) for breakdown in solids
has been used also for charge generation in liquids~\citep{Halpern1969a,Jadidian2013}.
Photoionization could also have an important role
in the generation of new charges~\citep{Lundgaard1998,Lesaint2016},
and adding field ionization and photoionization could improve the model.
In addition,
when ionizing neutral molecules,
the field-dependent ionization potential~\citep{Smalo2011}
should also be taken into account.
This kind of additions add complexity to the model,
but Monte Carlo (MC)~\citep{Metropolis1949} methods can aid
in keeping the added computational cost low.
There are also some parts of the current model where
MC could be reasonable to use.
For instance,
for electron detachment from an anion
and for avalanche growth from a single electron,
since a large number of electrons is needed
to model an avalanche through the average growth $\alpha$.

The degree of branching is lower than desired,
with more or less only one major branch,
and thus the simulations resemble more the 3rd mode
or the start of the 4th mode
than the 2nd mode of a streamer.
It is worthwhile noting that streamers branch far less
in cyclohexane than in mineral oil,
but the addition of low-IP additives
increases the branching~\citep{Lesaint2000}.
The shapes of the simulated streamers
do resemble the shape of streamers in longer gaps~\citep{Lesaint2000},
however,
while including additives in the model increases the propagation speed,
the degree of branching is not increased.
Although branching is thought of
as a mechanism for regulating the propagation speed,
it could be the other way around.
With nothing to hold it back,
the foremost head should have the strongest electric field
and the fastest propagation.
If something is regulating the speed or field
of the foremost head,
however,
then other heads are given a better chance of propagation,
increasing the number of branches,
which in turn may regulate
the electric field
of all the branches.
In the present model,
there is nothing holding the foremost head back,
since the only time scale included is that of the electron avalanches.
If,
for instance,
the time required for bubble nucleation
or the time for charges to move through the streamer structure
(streamer dynamics)
is important,
it may result in a disadvantage for the foremost head.
This is, however, not included,
and the potential of each streamer head is
instantly updated each simulation step.
The shape chosen for the streamer heads
could also be a major reason for the low degree of branching.
For a hyperboloid,
the electric field declines as $r^{-1}$ in front,
and the high-field region extends much further in the front
than on the sides.
Conversely,
the field from a monopole declines like $r^{-2}$
in all directions,
and could as such facilitate branching.
In such a model,
however,
the high field would be in a region closer to the streamer heads,
making an SCLF approach even more relevant.

The simplicity of the presented model
comes with several limitations,
as discussed above,
however,
a simple model is also a good place to start.
It makes it possible to identify
whether a certain mechanism is important or not
at a relatively low computational cost.
Consider \cref{fig:zapsim_stat_v_ct_sst(c)_mod},
which shows that the computational time for
breakdown streamers averages to about one hour,
using a single core on a regular desktop computer.
The simulation time is of course strongly dependent on
the number of seeds,
streamer heads,
and simulation steps,
but with such a low base case,
it is possible to perform a lot of simulations
to gather statistics
on a normal desktop computer.
Contrary to lattice models,
the presented model
is based on physical processes,
and the results are thus easier to evaluate.
FEM models may be better in the end,
but for now, such models cannot model a complete breakdown.
They are also simplified,
for example
in the sense that phase changes
are not accounted for~\citep{Jadidian2014}.
Both lattice and FEM models
demands much computational power
and the mesh size becomes an important parameter,
however,
this is avoided in the model presented.
Instead of dealing with processes at discrete point
or in discretized elements,
the model deals with discrete points that move.
This approach makes sense when considering
charge generated by electron avalanches
at some distance from the streamer structure,
or a streamer moving in discrete steps.
For details on processes inside
or very close to the streamer,
however,
a FEM approach seems more reasonable,
and could provide valuable input
to models on a larger scale.

}

%


\section{Conclusion}\label{sec:conclusion}{

A simple simulation model for streamer propagation
has been presented.
The streamer is represented by a collection
of hyperbolic streamer heads,
and is responsible for propagating the electric field
from the needle electrode.
In high-field areas,
electrons detach from ions present in the liquid,
and may turn into avalanches.
If an avalanche meets the Townsend--Meek criterion,
a new streamer head is added at its position,
causing the streamer to propagate.
As demonstrated,
the model has some limitations,
the inception voltage is too high
while the degree of branching is low.
These issues are discussed and explained,
and directions for a systematic way
of further developments are described.
The main feature missing in the model
is a proper representation of the dynamics of the streamer channel,
however,
the charge generation
and the electric field calculation
can be improved as well.
The approach to streamer propagation applied here
is different from that used by other models.
The principle behind the model is simple,
it is founded on physical mechanisms,
and provides interesting information
about how an avalanche-driven breakdown may occur.
The simple model has its advantages
in that it can be used to identify important mechanisms,
without demanding excessive computational power.

}

%


\section*{Acknowledgements}
The authors would like to thank
Lars Lundgaard and Dag Linhjell
for interesting discussions and
for sharing their experience about experiments.

This work has been supported by
The Research Council of Norway (RCN),
ABB and Statnett,
under the RCN contract 228850.


%

\appendix

\section{Prolate Spheroid Coordinates}{\label{sec:pscoords}

\begin{figure}[!b]
    \centering
    \includegraphics[width=0.85\linewidth]{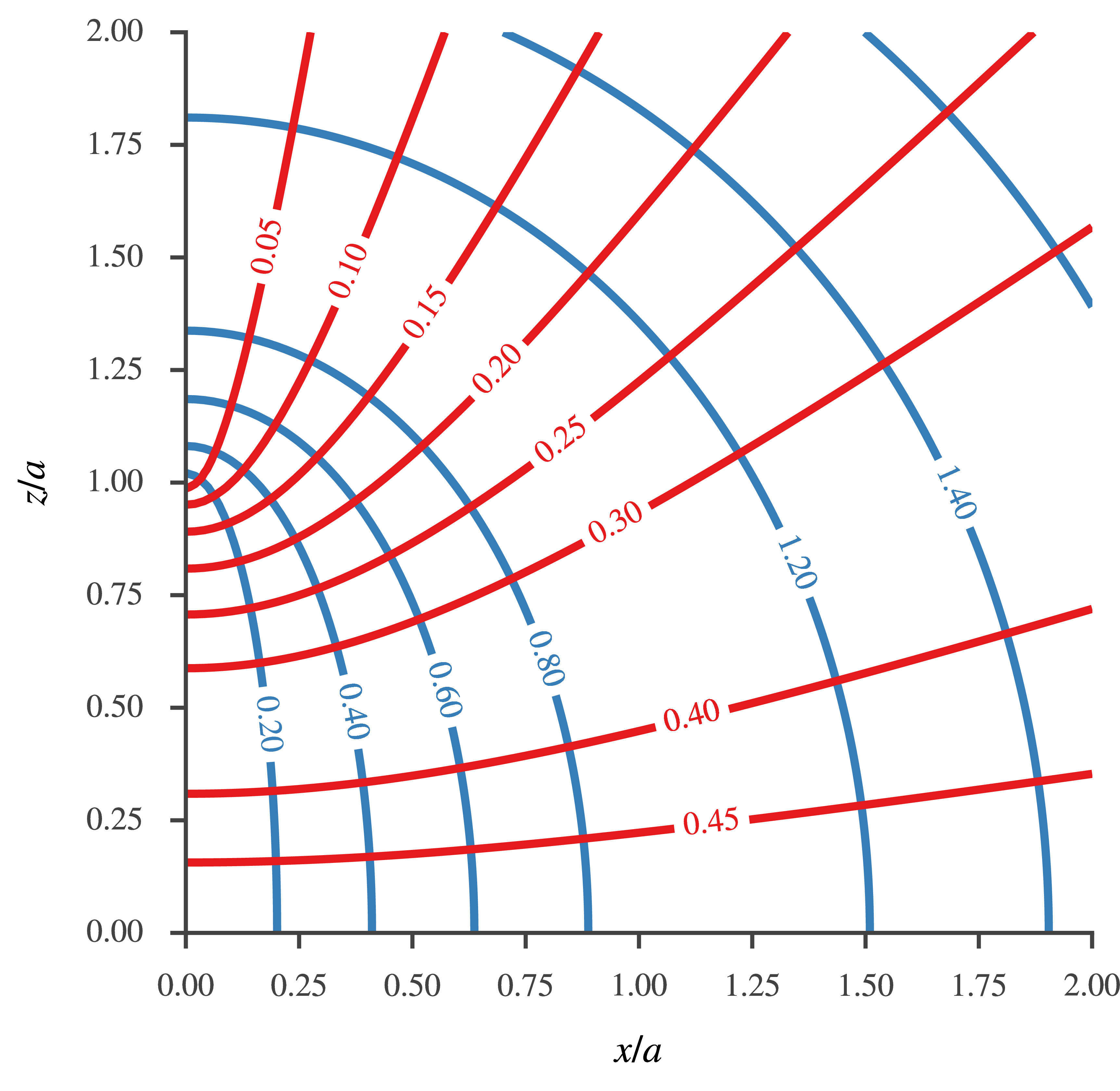}
    \caption{
        In prolate spheroid coordinates,
        spheroids (blue) are given by a constant $\mu$,
        and hyperboloids (red) have a constant $\nu$.
        Here, $\nu$ is given in units of $\pi$.}
    \label{fig:ps_coords}
\end{figure}

\begin{figure}[!b]
    \centering
    \includegraphics[width=0.95\linewidth]{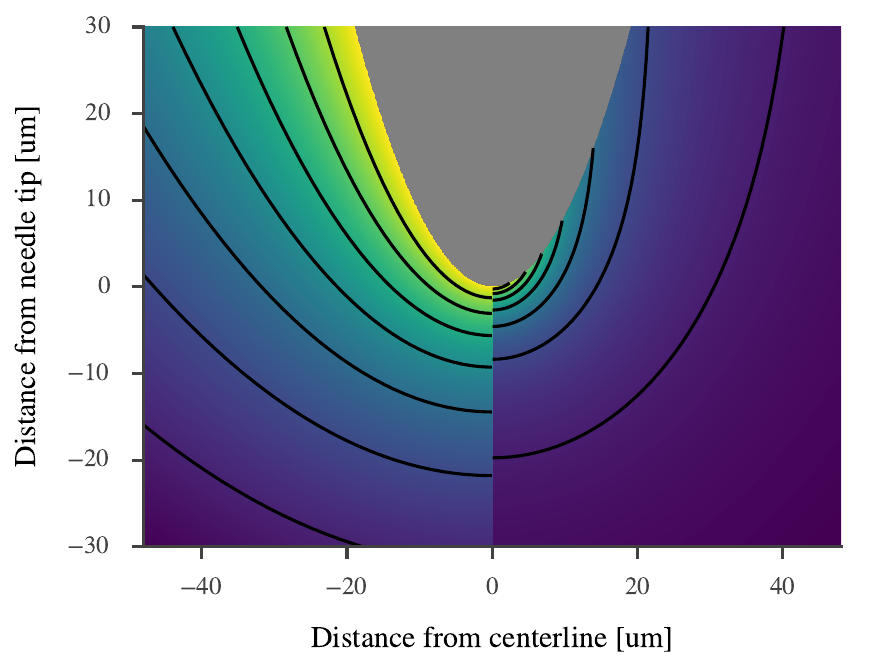}
    \caption{
        Electric potential (left) and electric field strength (right),
        for a region close to a needle (center, gray)
        placed \SI{10}{\milli\metre} from a grounded plane.
        The contour lines give a qualitative impression
        of how the respective magnitudes change as a function of position.
        A linear scale is used for both sides,
        and the magnitudes are linearly dependent on the potential of the needle.
        }
    \label{fig:e_str}
\end{figure}

Prolate spheroid coordinates involves
a set of hyperbolas and ellipsoids
revolved around the center axis,
forming hyperboloids and prolate spheroids.
The two focal points,
of the hyperbolas as well as ellipsoids,
are located at a distance $a$ from the plane.
The hyperbolic coordinate is $\mu \in [0,\infty\rangle$,
the elliptic coordinate is $\nu \in [0,\pi]$,
and rotation about the center is given by $\phi\in [0,2\pi]$.
The definition used here is
\begin{align}
    x &= a \, \sinh \mu \, \sin \nu \, \cos \phi \,, \label{eq:ps2x} \\[0.25 em]
    y &= a \, \sinh \mu \, \sin \nu \, \sin \phi \,, \label{eq:ps2y} \\[0.25 em]
    z &= a \, \cosh \mu \, \cos \nu \,. \label{eq:ps2z}
\end{align}
\Cref{fig:ps_coords} illustrates the coordinate system,
where a constant $\mu$ gives a prolate spheroid,
\begin{align}
    \frac{z^2      }{a^2 \cosh^2 \mu} +
    \frac{x^2 + y^2}{a^2 \sinh^2 \mu}
    = 1 \,,
\end{align}
and a constant $\nu$ gives a hyperbola,
\begin{align}
    \frac{z^2      }{a^2 \cos^2 \nu} -
    \frac{x^2 + y^2}{a^2 \sin^2 \nu}
    = 1 \,.
\end{align}
Transformation from Cartesian
to prolate spheroid coordinates is obtained through
\begin{align}
    2a \cosh \mu &= p + m \,, \label{eq:cosh_mu} \\[0.25 em]
    2a \cos \nu  &= p - m \,, \label{eq:cos_nu}  \\[0.25 em]
    \tan \phi &= y/x \,,        \label{eq:tan_phi}
\end{align}
where
\begin{align}
    p   &= \sqrt{ x^2 + y^2 + (z + a)^2 } \,, \\[0.25 em]
    m   &= \sqrt{ x^2 + y^2 + (z - a)^2 } \,,
\end{align}
and are the distances between a given point
and the two focal points.
Prolate spheroid coordinates exists in many forms.
In some cases,
it is easier to work with substitutions
such as $\xi = \sin \nu$,
however,
starting with trigonometric functions
allows for greater flexibility
through relations such as
$\sin^2 + \cos^2 = 1$.

Scale factors $h$ are useful to define
when transforming between coordinate systems.
The scale factor for $\nu$, for instance, is found from
\begin{equation}
    h_\nu = \frac{\d l}{\d \nu}
          = \sqrt{
                \left( \frac{\d x}{\d \nu} \right)^2 +
                \left( \frac{\d y}{\d \nu} \right)^2 +
                \left( \frac{\d z}{\d \nu} \right)^2 }
                \,.
                \label{eq:h_nu_def}
\end{equation}
Solving this,
and the similar expressions for the other coordinates,
yields
\begin{align}
    h_\nu = h_\mu &= a \, \sqrt{\sinh^2\mu + \sin^2\nu} \,, \label{eq:h_nu} \\[0.25 em]
    h_\phi        &= a \, \sinh\mu \, \sin\nu \,. \label{eq:h_phi}
\end{align}
These are useful when defining the spatial derivative,
\begin{equation}
    \vec{\nabla} =
                   \frac{\hat{\mu}}{h_\mu} \partial_\mu
                 + \frac{\hat{\nu}}{h_\nu} \partial_\nu
                 + \frac{\hat{\phi}}{h_\phi} \partial_\phi
                 \,. \label{eq:ps_nabla}
\end{equation}
The electric potential $V$
and the electric field $\vec{E}$
are found by solving
the Laplace equation, ${\vec{\nabla}^2 V = 0}$.
For a system where the hyperboloids represent equipotential surfaces,
$V = V(\nu)$,
the Laplace equation is satisfied for~\citep{Moon1971}
\begin{equation}
    V (\nu) = A + C \ln \tan \frac{\nu}{2} \,,
    \label{eq:ps_V}
\end{equation}
where the constants $A$ and $C$ are defined by boundary conditions.
Given 
${V(\nu = \pi/2) = 0}$ at the $xz$-plane
and ${V(\nu = \nu_0) = V_0}$ at the $\nu_0$-hyperboloid,
yields $A = 0$ and
\begin{equation}
    C = \frac{V_0}{\ln \tan (\nu_0 / 2)} \,.
    \label{eq:ps_C}
\end{equation}
Consequently, the electric field
{$\vec{E} = - \vec{\nabla} V$} becomes
\begin{equation}
    \vec{E}
        = \frac{C \, \hat{\nu}}{h_\nu \sin \nu}  \,,
    \label{eq:ps_E}
\end{equation}
where
$\hat{\nu}$ is unit length in the direction of $\vec{\nu}$,
\begin{equation}
    \hat{\nu}
        = \frac{\vec{\nu}}{\abs{\nu}}
        = \frac{\partial_\nu \left( \vec{x} + \vec{y} + \vec{z} \right)}{h_\nu}
        = \frac{ \vec{x} + \vec{y} - \vec{z} \tan^2\nu }{h_\nu \tan\nu}
        \,. \label{eq:ps_nu_hat}
\end{equation}
\Cref{eq:ps_V,eq:ps_E} are both illustrated in \cref{fig:e_str}.
The figure shows the differences in behavior
between the electric potential and the electric field,
the latter increases rapidly close to the tip of the hyperboloid.
%

%
Explicit transformation between Cartesian
and prolate spheroid coordinates
requires trigonometric and hyperbolic functions,
which are costly when it comes to computations.
There is, however, no need to calculate
$\mu$, $\nu$, and $\phi$ explicitly,
as both the potential \cref{eq:ps_V}
and the electric field \cref{eq:ps_E}
may be obtained by using
\cref{eq:cosh_mu,eq:cos_nu}, and trigonometric relations
such as
\begin{align}
    2a\sin \nu &= 2a\sqrt{1 - \cos^2 \nu} \notag\\
               &= \sqrt{4a^2 - ( p - m)^2} \,.
\end{align}

}

%


\end{document}